%% file: ms.tex
\documentclass{vldb}
\pdfoutput=1

\usepackage{times}
\usepackage{graphicx}
\usepackage[labelformat=simple]{subcaption}

\usepackage{amsmath}
\usepackage{ wasysym }
\usepackage[export]{adjustbox}
\usepackage{url}
\usepackage{balance}  
\newtheorem{exmp}{Example}[section]
\newtheorem{definition}{Definition}
\usepackage{floatrow}
\usepackage[normalem]{ulem}
\newfloatcommand{capbtabbox}{table}[][\FBwidth]

\usepackage[linesnumbered,ruled,vlined]{algorithm2e}
\SetKwProg{Fn}{Function}{ is}{end}
\SetAlFnt{\scriptsize}

\def\BibTeX{{\rm B\kern-.05em{\sc i\kern-.025em b}\kern-.08em
    T\kern-.1667em\lower.7ex\hbox{E}\kern-.125emX}}

\vldbTitle{Distributed Dependency Discovery}
\vldbAuthors{Hemant Saxena, Lukasz Golab, Ihab F. Ilyas}
\vldbDOI{https://doi.org/10.14778/xxxxxxx.xxxxxxx}
\vldbVolume{12}
\vldbNumber{xxx}
\vldbYear{2018}

\makeatletter
\newcommand{\removelatexerror}{\let\@latex@error\@gobble}
\def\@copyrightspace{\relax}
\def\@mkbibcitation{\relax}
\makeatother

\newcommand{\myalgorithma}{
\begingroup
\removelatexerror
\begin{algorithm}[H]
    \DontPrintSemicolon
  \Fn{TANE (Relation $r$, Schema $R$)}{
      \For {$A \in R$} {
      	$L_1$ = $L_1$ $\cup$ \{$(A, genEQClass(A, r))$\} \\
      }
      $l$ = 1 \\
      
      \While{$L_l$ $\neq$ $\phi$}{
          computeDependencies($L_{l}$) \\
          prune($L_{l}$) \\
          generateNextLevel($L_l$) \\
          $l$++ \\
      }   
  }
  \Fn{computeDependencies (Level $L_{l}$)}{
      \For{$X \in L_{l}$}{
          \For{$A \in X$}{
              checkRefinement($X/A$, $X$, ($|\pi_{X/A}|$, $|\pi_{X}|$))
          }
      }   
  }
  \Fn{generateNextLevel (Level $L_l$)}{
      \For{($X, \pi_X, Y, \pi_Y$) $\in$ join$(L_l, L_l)$}{
              $\pi_{X \cup Y}$ = genEQClass($X \cup Y$, ($\pi_X$, $\pi_Y$)) \\
              $L_{l+1}$ = $L_{l+1}$ $\cup$ \{($X \cup Y$, $\pi_{X \cup Y}$)\}
      }   
  }
  \end{algorithm}
\endgroup
  }
  
  \newcommand{\myalgorithmb}{
\begingroup
\removelatexerror
  \begin{algorithm}[H]
    \DontPrintSemicolon
  \Fn{TANE (Relation $r$, Schema $R$)}{
      \For {$A \in R$} {
      	\textcolor{blue}{$L_1$ = $L_1$ $\cup$ $\{(A, |genEQClass(A, r)|)$\}}\\
      }
      $l$ = 1 \\
      
      \While{$L_l$ $\neq$ $\phi$}{
          computeDependencies($L_{l}$) \\
          prune($L_{l}$) \\
          generateNextLevel($L_l$) \\
          $l$++ \\
      }   
  }
  \Fn{computeDependencies (Level $L_{l}$)}{
      \For{$X \in L_{l}$}{
          \For{$A \in X$}{
              checkRefinement($X/A$, $X$, ($|\pi_{X/A}|$, $|\pi_{X}|$))
          }
      }   
  }
  \Fn{generateNextLevel (Level $L_l$)}{
      \For{($X, \pi_X, Y, \pi_Y) \in$ join$(L_l, L_l)$}{
              \textcolor{blue}{$\pi_{X \cup Y}$ = genEQClass($X \cup Y$, r)} \\
              \textcolor{blue}{$L_{l+1}$ = $L_{l+1}$ $\cup$ \{($X \cup Y$, $|\pi_{X \cup Y}|$)\}} \\
      }   
  }
  \end{algorithm}
\endgroup
  }

\newcommand{\myalgorithmc}{
\begingroup
\removelatexerror
\begin{algorithm}[H]
  \DontPrintSemicolon
  \Fn{fastFD (Relation $r$, Schema $R$)}{
      
      $EV_I = \{\}$ \\
      
      generateEvidence($r$, $R$) \\
      $EV_I$ = sort($EV_I$) \\
      $FDs$ = setCover($EV_I$) \\
  }
  \Fn{generateEvidence (Relation $r$, Schema $R$)}{
  		$EQ = \{\}$ \\
  		\For{$A$ $\in$ $R$}{
          $EQ$ = $EQ$ $\cup$ genEQClass($A, r$) \\
      }
      \For{$\pi \in EQ$}{
      \For {($t_i$, $t_j$) $\in$ join($\pi$,$\pi$)} {
            $EV_I$ = $EV_I$ $\cup$ genEVSet($t_i$,$t_j$, $\{\neq\}$)
       }
      }   
  }
  \end{algorithm}
\endgroup
  }
 
  \newcommand{\myalgorithmd}{
\begingroup
\removelatexerror
  \begin{algorithm}[H]
  \DontPrintSemicolon
  \Fn{fastFD (Relation $r$, Schema $R$)}{
      $EV_I = \{\}$ \\
      \textcolor{blue}{generateEvidence($r$)} \\
      $EV_I$ = sort($EV_I$) \\
      $FDs$ = setCover($EV_I$) \\
  }
  \Fn{generateEvidence (Relation $r$)}{
      \For {\textcolor{blue}{($t_i$, $t_j$) $\in$ join($r$,$r$)}} {
            $EV_I$ = $EV_I$ $\cup$ genEVSet($t_i$,$t_j$, $\{\neq\}$)
       }
  }
  \end{algorithm}
\endgroup
  }

\begin{document}

\title{Distributed Dependency Discovery}

\numberofauthors{3} 
\author{
\alignauthor
Hemant Saxena\\
      \affaddr{University of Waterloo}\\
      \email{h2saxena@uwaterloo.ca}
\alignauthor Lukasz Golab\\
      \affaddr{University of Waterloo}\\
      \email{lgolab@uwaterloo.ca}
\alignauthor Ihab F. Ilyas\\
      \affaddr{University of Waterloo}\\
      \email{ilyas@uwaterloo.ca}
}

\maketitle
\sloppy


\begin{abstract}
We analyze the problem of discovering dependencies from distributed big data.  Existing (non-distributed) algorithms focus on minimizing computation by pruning the search space of possible dependencies.  However, distributed algorithms must also optimize communication costs, especially in shared-nothing settings, leading to a more complex optimization space. To understand this space, we introduce six primitives shared by existing dependency discovery algorithms, 
corresponding to data processing steps separated by communication barriers.  Through case studies, we show how the primitives allow us to analyze the design space and develop communication-optimized implementations.
Finally, we support our analysis with an experimental evaluation on real datasets.

\end{abstract}

\section{Introduction}
\label{sec:intro}
\input{introduction.tex}
\input{background.tex}

\section{Primitives for Dependency Discovery} \label{sec:primitives}
\input{primitives.tex}
\section{Case study 1: TANE}
\label{sec:tane}
\input{tane.tex}
\section{Case study 2: FastFDs}
\label{sec:fastfd}
\input{fastfd.tex}

\section{Case study 3: HyFD}
\label{sec:hyfd}
\input{hyfd.tex}

\section{Experiments} \label{sec:eval}
\input{evaluation}

\section{Conclusions} \label{sec:conclusions}


In this paper, we took a first step towards understanding the problem of distributed dependency discovery. We proposed an analysis framework consisting of six primitives that correspond to the data processing steps of existing discovery algorithms for UCCs, FDs, ODs and DCs. The primitives allowed us to analyze the algorithms in terms of their communication and computation costs and enabled an exploration of the space of possible optimizations. We demonstrated this exploration via case studies and an empirical evaluation. In particular, our experimental results showed that the execution plans which revisit the design decisions made in the original non-distributed algorithms outperform the straightforward distributed plans. 
This implies that new optimizations are required even for existing algorithms to handle distributed datasets.

A natural direction for future work is to build a cost-based optimizer for dependency discovery based on our primitives. Given a dependency that is to be discovered, a particular dataset and a system configuration, the goal of the optimizer will be to select the best physical execution plan.

\bibliographystyle{abbrv}
\bibliography{ms}

\end{document}

%% file: introduction.tex
Column dependencies such as candidate keys or Unique Column Combinations (UCCs), Functional Dependencies (FDs), Order Dependencies (ODs) and Denial Constraints (DCs) are critical in many data management tasks including schema design, data cleaning, data analytics and query optimization.  Despite their importance, dependencies are not always specified in practice, and even if they are, they may change over time. Furthermore, dependencies that hold on individual datasets may not hold after performing data integration.  As a result, there has been a great deal of research on automated discovery of dependencies from data; see, e.g., \cite{profiling,Liu:2012:DDD:2122270.2122559} for recent surveys. 

Existing work on dependency discovery proposes methods for pruning the exponential search space in order to minimize computation costs. However, existing methods assume a centralized setting where the data are stored  locally.
In contrast to centralized settings, in modern big data infrastructure, data are naturally partitioned (e.g., on HDFS \cite{hdfs}) and computation is parallelized (e.g., using Spark \cite{spark}) across multiple compute nodes.  In these cases, it is inefficient at best and infeasible at worst to move the data to a centralized profiling system, motivating the need for
distributed profiling.

\begin{figure}[t]
    \begin{subfigure}[t]{0.45\textwidth}
        \centering
        \includegraphics[scale=0.27,valign=t]{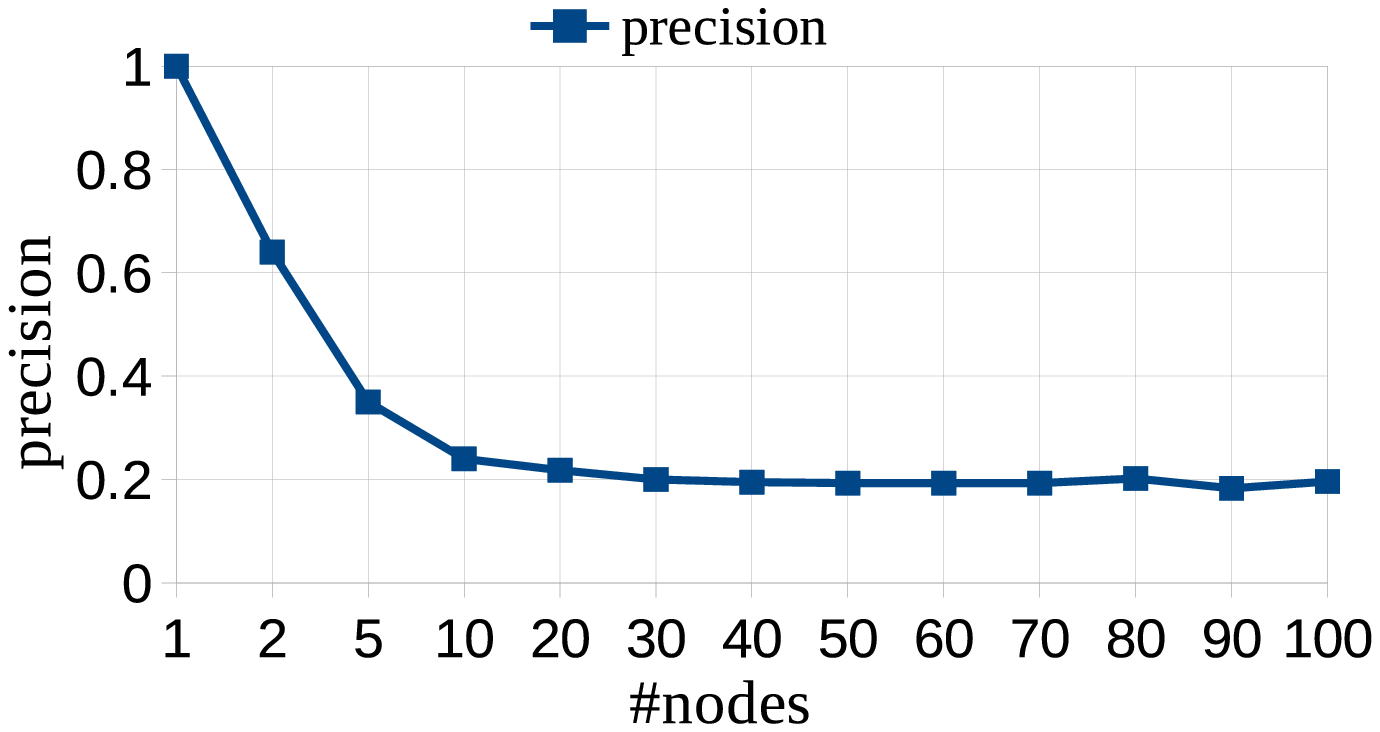}
        \caption{Drop in precision (i.e., the fraction of locally discovered FDs that hold over the entire dataset) with number of nodes}
    \label{fig:fdOnParts}
    \end{subfigure}%
~
\begin{subfigure}[t]{0.45\textwidth}
        \centering
        \includegraphics[scale=0.28,valign=t]{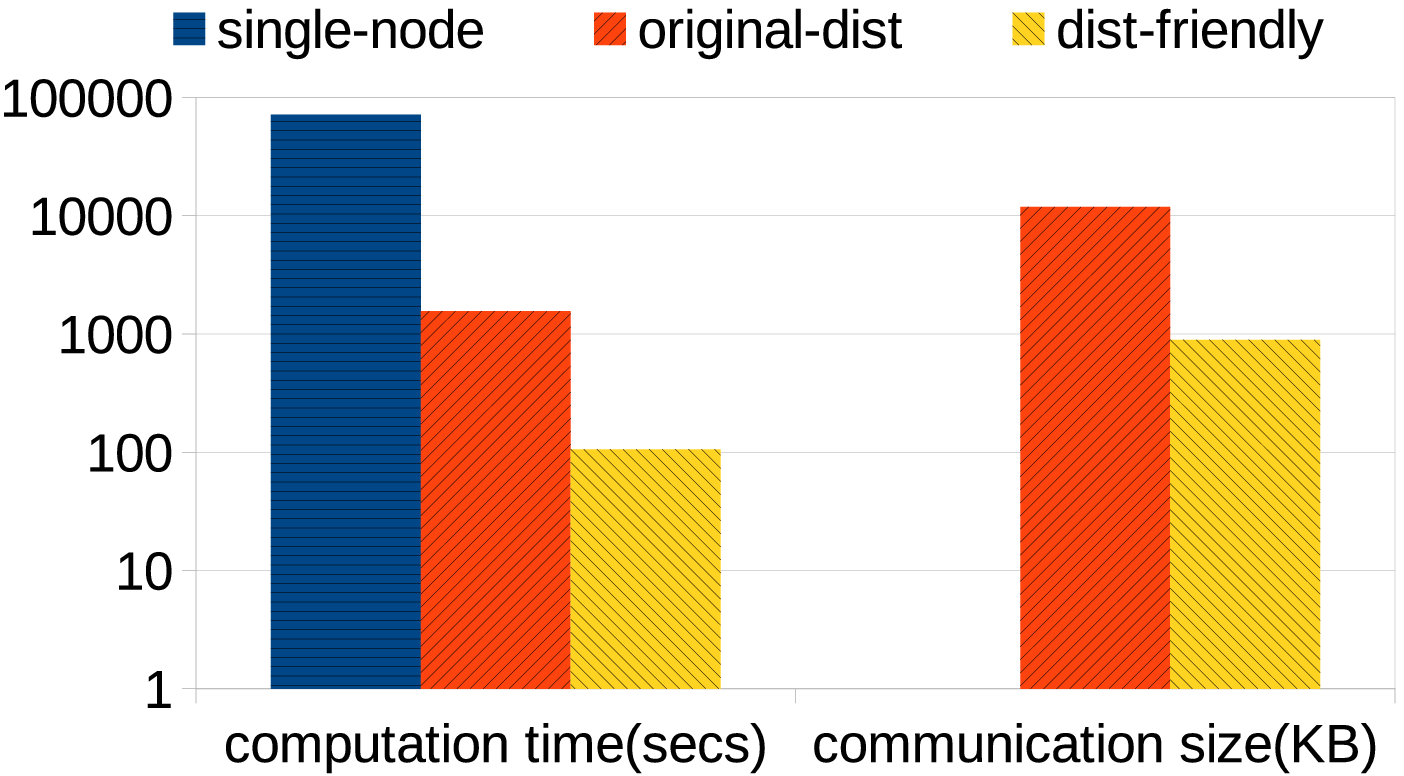}
        \caption{Computation and communication costs of FastFDs}
    	\label{fig:intro-plot}
    \end{subfigure}
\caption{Precision of naive solution for distributed FD discovery and performance of three FastFDs implementations}
\end{figure}

In distributed environments, ensuring good performance requires minimizing computation and communication costs.
 A na\"ive solution to minimize communication costs is to allow no data communication at all: each node locally discovers dependencies from the data it stores, and then we take the intersection of the locally-discovered dependencies.  To see why this approach fails, consider a table with a schema $(A,B)$ and assume the table is partitioned across two nodes: the first node storing tuples $(a_1, b_1)$, $(a_1, b_1)$, and the second node storing tuples $(a_1, b_2)$, $(a_1, b_2)$. The FD $A \rightarrow B$ locally holds on both nodes but it does not hold globally over the whole table. We show in Figure~\ref{fig:fdOnParts} that this problem gets worse quickly, even for as few as ten nodes, where the majority of the dependencies discovered locally do not hold on the entire dataset (TPC-H \textit{lineitem} table with one million rows \cite{tpch}). Notably, discovering dependencies from a sample has a similar problem.
 

Another possible solution is to parallelize existing non-distributed dependency discovery algorithms in a straightforward way. The problem with this approach is that existing algorithms often generate large intermediate results to minimize computation, leading to high communication overhead. Other problems include parallelizing the computation and load balancing---issues that, naturally, were not considered in centralized implementations.

In this paper, we argue that to implement efficient distributed algorithms, an end-user needs to (i) systematically analyze the space of possible optimizations, i.e., identify the core data processing steps and optimize for both computation and communication when parallelizing these steps, and (ii) tune the physical implementations of these steps, i.e.,  design a distribution strategy for a given workload and the available computational and memory resources.
To facilitate the end-user in overcoming these challenges, we decompose existing dependency discovery algorithms into six logical \emph{primitives}, corresponding to data processing steps separated by communication barriers. The primitives allow us to rewrite the algorithms, analyze the computation and communication costs of each step, and explore the space of possible distributed designs, each with different performance characteristics.

From the point of view of an end-user, our primitive-oriented framework decouples writing distributed versions of the algorithms from tuning their physical implementations. We refer to the logical rewrites using our primitives as \textit{logical discovery plans} or LDPs, and their physical implementations as \textit{physical discovery plans} or PDPs.  We present case studies (Sections \ref{sec:tane} through \ref{sec:hyfd}), showing how our primitives allow us to explore the space of possible designs. In particular, for each algorithm, we write two LDPs using our primitives. One LDP follows the design principles of the original non-distributed implementations, and the other LDP modifies the original algorithm to make it distribution-friendly. We then demonstrate that different physical implementations of the primitives lead to different PDPs for the same LDP.


Figure~\ref{fig:intro-plot} illustrates the impact of exploring different design options using our primitives.  We compare the performance of three versions of the FastFDs algorithm \cite{fastfd} for FD discovery on a \textit{homicide} data set with 100,000 rows and 24 columns (details in Section~\ref{sec:eval}). The first version, \textit{single-node}, is the original FastFDs algorithm executed on a single machine.  The second, \textit{original-dist} is a distributed version of the original algorithm running on a 55-worker Spark cluster, which is faster than the single-machine solution but has a significant communication overhead.  The third, \textit{dist-friendly}, is a distribution friendly version that reduces the data communication costs and is over an order of magnitude faster than \textit{original-dist}.

To summarize, our contributions are as follows.
\begin{enumerate}
\item
We propose a generalized framework for analyzing dependency discovery algorithms (including UCCs, FDs, ODs and DCs), which consists of six primitives that serve as building blocks of existing algorithms.
\item
Using case studies, we illustrate how the primitives allow us to (i) decouple logical designs from physical implementations, and (ii) analyze the cost of individual data processing steps. 
\item
Using the proposed primitives, we implement and experimentally evaluate different distributed versions of seven existing dependency discovery algorithms on several real datasets.


\end{enumerate}


\textbf{Prior Work:}
There has been recent work on parallelizing dependency discovery algorithms across multiple threads, but it considers a singe-node shared-everything architecture where communication costs are not a bottleneck \cite{paMiDe}.  There is also some early work on distributed FD discovery.  However, it suffers from the same issues as the na\"ive solution we mentioned earlier (i.e., it returns locally-discovered FDs which may not hold globally) \cite{approxFDDist}, or it assumes that data are partitioned vertically and ensures efficiency by limiting the search space to FDs with single attributes \cite{vertDist}.
Finally, in our recent work \cite{distribute-fastfds}, we proposed a distributed FastFDs implementation, which is one of the designs we analyze and express using primitives in this paper (corresponding to Section~\ref{sec:imp-fastfd-phy1}). 

\textbf{Roadmap:}
Section~\ref{sec:preliminaries} explains existing dependency discovery algorithms,
Section~\ref{sec:primitives} presents the primitives,
Sections~\ref{sec:tane} through \ref{sec:hyfd} present cases studies in which we analyze FD discovery algorithms using the primitives,
Section~\ref{sec:eval} presents experimental results,
and Section~\ref{sec:conclusions} concludes the paper.

%% file: background.tex
\section{Preliminaries} \label{sec:preliminaries}
We begin by defining the fundamental concepts and data structures used in dependency discovery algorithms.  Table~\ref{tab:symbols} lists the important symbols used in this paper.

\begin{table}[t]
\caption{Summary of symbols}
\centering
\resizebox{0.8\textwidth}{!}{
\label{tab:symbols}
\begin{tabular}{ |c|c| } 
 \hline
 \textbf{Symbol} & \textbf{Meaning} \\\hline
 $R$ & A relation \\ \hline
 $r$ & An instance of $R$ \\ \hline
 $n$ & Number of tuples in $r$ \\ \hline
 $m$ & Number of attributes $R$ \\ \hline
 $\pi_X$ & Equivalence classes of $X \subseteq R$ \\ \hline
 $EV(t_i,t_j)$ & Evidence set due to tuple pair ($t_i$, $t_j$) \\ \hline
 $k$ & Number of workers/compute nodes \\ \hline
 \textbf{Y} & Maximum computation done by any worker \\ \hline
 \textbf{X} & Maximum data sent to any worker \\ \hline
\end{tabular}
}
\end{table}

\begin{table}[t]
\centering
\caption{Tax data records}
\resizebox{\textwidth}{!}{
\begin{tabular}{ |c|c|c|c|c|c|c|c|c|c|c| } 
 \hline
 tid & ID & GD & AC & PH & CT & ST & ZIP & SAL & TR & STX \\\hline \hline
 t1 & 1009 & M & 304 & 232-7667 & Anthony & WV & 25813 & 5000 & 3 & 2000 \\\hline
 t2 & 2136 & M & 719 & 154-4816 & Denver & CO & 80290 & 60000 & 4.63 & 0 \\\hline
 t3 & 0457 & F & 636 & 604-2692 & Cyrene & MO & 64739 & 40000 & 6 & 0 \\\hline
 t4 & 1942 & F & 501 & 378-7304 & West Crossett & AR & 72045 & 85000 & 7.22 & 0 \\\hline
 t5 & 2247 & M & 319 & 150-3642 & Gifford & IA & 52404 & 15000 & 2.48 & 50 \\\hline
 t6 & 6160 & M & 970 & 190-3324 & Denver & CO & 80251 & 60000 & 4.63 & 0 \\\hline
 t7 & 4312 & F & 501 & 154-4816 & Kremlin & AR & 72045 & 70000 & 7 & 0 \\\hline
 t8 & 3339 & F & 304 & 540-4707 & Kyle & WV & 25813 & 10000 & 4 & 500 \\\hline
\end{tabular}
}
\label{tab:small}
\end{table}

\subsection{Definitions}
\label{sec:defs}
Let $R = \{A_1, A_2, ..., A_m\}$ be a set of attributes describing the schema of a relation $R$ and let $r$ be a finite instance of $R$ with $n$ tuples.
\begin{definition}
\textbf{Minimal unique column combination:}
An attribute combination $X \subseteq R$ is a unique column combination (UCC) if $X$ uniquely identifies tuples in $r$, i.e., if no two tuples in $r$ have the same values of $X$.  $X$ is a minimal UCC if no proper subset of it is a UCC.
\end{definition}
\begin{definition}
\textbf{Minimal functional dependency:}
Let $X \subset R$ and $A \in R$. A functional dependency (FD) $X \rightarrow A$ holds on $r$ iff for every pair of tuples $t_i, t_j \in r$ the following is true: if $t_i[X] = t_j[X]$, then $t_i[A] = t_j[A]$.
An FD $X \rightarrow A$ is minimal if $A$ is not functionally dependent on any proper subset of $X$.
\end{definition}
\begin{definition}
\textbf{Minimal order dependency:}
Let $X \subset R$ and $A \in R$.  An order dependency (OD) $X \longmapsto Y$ holds if sorting $r$ by $X$ means that $r$ is also sorted by $Y$. An OD $X \longmapsto A$ is minimal if $A$ is not order dependent on any proper subset of $X$.
\end{definition}
\begin{definition}
\label{def:dc}
\textbf{Denial constraint:}
A Denial constraint (DC) $\psi$ is a statement of the form 
$\psi: \forall t_i, t_j \in r, \neg (P_1 \wedge ... \wedge P_k)$
where $P_i$ is of the form $v_1 \phi v_2$ with $v_1, v_2 \in t_x[A]$, $x \in \{i, j\}$, $A \in R$ and $\phi \in \{=, \neq, <, \leq, >, \geq\}$.
The expression inside the brackets is a conjunction of predicates, each containing two attributes from $R$ and an operator $\phi$.
An instance $r$ satisfies $\psi$ $iff$ $\psi$ is satisfied for any two tuples $t_i$, $t_j$ $\in r$.
\end{definition}

\begin{exmp}
Consider the tax dataset in Table~\ref{tab:small}.
The set $\{AC, PH\}$ is a UCC.  Two persons with same zip code live in the same state, therefore the FD $ZIP \rightarrow ST$ holds. The single tax exemption decreases as salary increases, therefore the OD $SAL \longmapsto STX$ holds. If two persons live in the same state, the one earning a lower salary has a lower tax rate, therefore the following DC holds: $\forall t_i, t_j \in R,$ $\neg (t_i.ST = t_j.ST \wedge t_i.SAL < t_j.SAL \wedge t_i.TR > t_j.TR)$.
\end{exmp}

In the remainder of this paper, discovering dependencies refers to discovering \emph{minimal} dependencies.


\begin{definition}
\label{def:EQ}
\textbf{Equivalence classes:}
The equivalence class of a tuple $t \in r$ with respect to an attribute set $X \subseteq R$ is denoted by $[t]_X = \{u \in r | \forall A \in X\  t[A] = u[A]\}$.
The set $\pi_X = \{ [t]_X | t \in r\}$ contains the equivalence classes of r under X.
\end{definition}

Note that $\pi_X$ is a \textit{partition} of $r$ such that each equivalence class corresponds to a unique value of $X$.
Let $|\pi_X|$ be the number of equivalence classes in $\pi_X$, i.e., the number of distinct values of $X$.

\begin{definition}
\label{def:EV}
\textbf{Evidence sets:}
For any two tuples $t_i$ and $t_j$, in $r$, their evidence set $EV(t_i,t_j)$ is the set of predicates satisfied by them, drawn from some predicate space $P$.
\end{definition}

For example, recall the predicate space considered by DCs from Definition~\ref{def:dc}, namely those with two attributes from $R$ and an operator from $\phi$.  In Table~\ref{tab:small}, tuples $t_2$ and $t_6$ give $EV(t_2, t_6) = \{t_2.ID \neq t_6.ID,$ $t_2.ID \leq t_6.ID,$  $t_2.ID < t_6.ID,$ $t_2.GD = t_6.GD,$ $t_2.CT = t_6.CT,$ $t_2.ST = t_6.ST,$ $...\}$. For FDs and UCCs, it suffices to consider a restricted space of predicates that identify which attribute values are different.  Here, $EV(t_2, t_6) = \{t_2.ID \neq t_6.ID,$ $t_2.AC \neq t_6.AC,$ $t_2.PH \neq t_6.PH,$ $t_2.ZIP \neq t_6.ZIP\}$.  As we will show in Section~\ref{sec:algos}, some algorithms use evidence sets to identify dependencies that do \emph{not} hold.

\begin{definition}
\label{def:PR}
\textbf{Partition refinement:}
Partition $\pi$ \textit{refines} partition $\pi'$ if every equivalence class in $\pi$ is a subset of some equivalence class of $\pi'$.
\end{definition}


\textbf{Communication and computation cost:} Suppose we have $k$ workers or compute nodes.  Let $X_i$ and $Y_i$ be the amount of data sent to the $i^{th}$ worker and the computation done by the $i^{th}$ worker, respectively \cite{xu}.  The runtime of a distributed algorithm depends on the runtime of the slowest worker.  Thus, we will compute the following quantities for each tested algorithm:
\begin{align*}
\textbf{X} = \max_{i \in [1,k]} X_i & & 
\textbf{Y} = \max_{i \in [1,k]} Y_i \\
\end{align*}

\subsection{Algorithms} \label{sec:algos}

Dependency discovery algorithms can be classified into three categories: schema-driven, data-driven and hybrid.  

\begin{figure}
\raggedright
    \begin{subfigure}[t]{0.3\textwidth}
        \centering
        \begin{tabular}{ |c|c|c|c| } 
        \hline
        A & B & C & D \\\hline
        a & a & a & a \\\hline
        b & b & a & a \\\hline
        a & c & a & c \\\hline
        b & c & d & e \\\hline
        \end{tabular}
        \caption{Relation instance}
		\label{fig:small2}
    \end{subfigure}%
    ~ 
    \begin{subfigure}[t]{0.5\textwidth}
        \includegraphics[scale=0.2,valign=c]{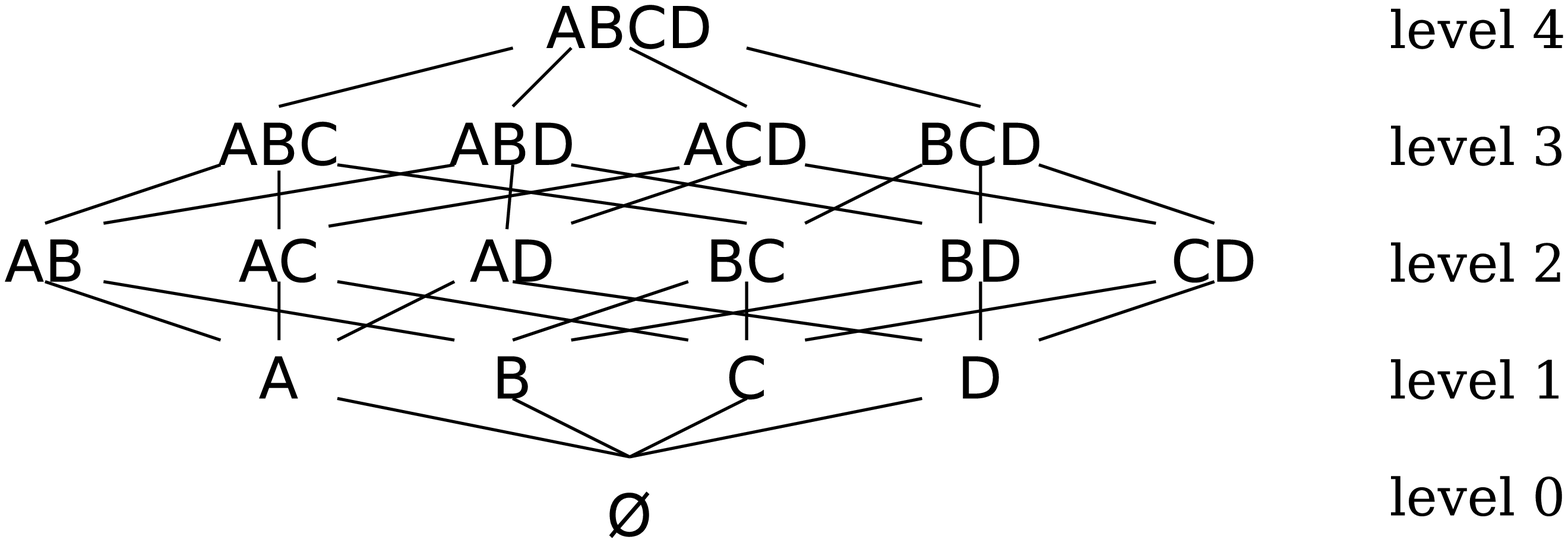}
        \caption{Attribute lattice}
        \label{fig:lattice}
    \end{subfigure}
    \caption{Example relation instance and attribute lattice}
\end{figure}

\textbf{Schema-driven algorithms} This class of algorithms traverses an \emph{attribute lattice} in a breadth-first manner, an example of which is shown in Figure~\ref{fig:lattice} for $R$ = $\{A, B, C, D\}$.  The nodes in the $i^{th}$ lattice level, denoted $L_i$, correspond to sets of $i$ attributes.  Each node also stores the equivalence classes (Definition~\ref{def:EQ}) corresponding to its attribute set.  Edges between nodes are based on a set containment relationship of their attribute sets.  The time complexity of schema-driven algorithms depends mainly on the size of the lattice, but not on the number of tuples. Therefore, these algorithms work well for large datasets with few columns.

Consider the TANE \cite{tane} algorithm for discovering FDs (FastOD \cite{fastod} is similar but it discovers ODs). For each lattice level, TANE performs three tasks: \textit{generate next level}, \textit{compute dependencies}, and \textit{prune}.  To generate the next level, TANE first creates new attribute sets by combining pairs of attribute sets from the current level; e.g., combining $AB$ and $AC$ gives $ABC$.  This corresponds to a self-join of the current level's attribute combinations.  Next, new equivalence classes are created by \emph{intersecting} pairs of equivalence classes from the current level.  For example, in  
Figure~\ref{fig:small2} we have $\pi_A$ $=$ $\{\{1, 3\},\{2, 4\}$, $\pi_{B}$ $=$ $\{\{1\},\{2\},\{3, 4\}\}$, $\pi_{C}$ $=$ $\{\{1, 2, 3\},$ $\{4\}\}$ and $\pi_{D}$ $=$ $\{\{1, 2\},\{3\},\{4\}\}$. Intersecting $\pi_{C}$ and $\pi_{D}$ gives $\pi_{\{C, D\}}$ $=$ $\{\{1, 2\}, \{3\}, \{4\}\}$.


Once the attribute sets and equivalence classes for the next level $L_l$ have been generated, the \textit{compute dependency} task discovers FDs of the form $X \setminus A \rightarrow A$ for all $X \in L_l$, and for all $A \in X$.  To determine if $X \setminus A \rightarrow A$, TANE checks if $\pi_{X\setminus A}$ \emph{refines}  $\pi_A$ (Definition~\ref{def:PR}). For example, in Figure~\ref{fig:small2}, $D \rightarrow C$ holds because $\pi_D$ refines $\pi_{CD}$; however, $C \rightarrow D$ does not hold because $\pi_C$ does not refine $\pi_{CD}$.

\textit{Compute dependencies} is simpler for FD and UCC discovery but more complex for ODs.  For FDs, it suffices to check if $|\pi_{X \setminus A}| = |\pi_{X}|$.  For UCCs, $X$ is a UCC if $|\pi_X|=r$, i.e., if all equivalence sets are singletons.  On the other hand, for an OD $X \longmapsto Y$ to hold, every set in $\pi_{X\setminus A}$ must be a subset of some set in $\pi_A$ and furthermore the elements must be ordered in the same way.

Finally, \emph{prune} leverages the fact that only minimal dependencies are of interest; for example, if $A \rightarrow D$ holds then $AB \rightarrow D$ is not minimal.  Depending on the algorithm, various pruning rules are applied to eliminate nodes from the lattice that cannot produce non-minimal dependencies. If a node is pruned, then any nodes connected to it can also be eliminated.  For example, for FDs, a node labelled with an attribute set $X$ can be pruned if $X$ is a key or $X\setminus A \rightarrow A$ was found to hold.  Returning to our example, $CD$ is pruned because $D \rightarrow C$ holds, and the following nodes are pruned because they correspond to keys: $AB$, $AD$, $BC$, and $BD$.

\textbf{Data-driven algorithms} This class of algorithms examines pairs of tuples to identify  evidence sets (Definition \ref{def:EV}) and violated dependencies; in the end, any dependencies not found to be violated must hold.  The time complexity of data-driven algorithms depends on the number of tuples, but not on the number of columns. Therefore, these algorithms tend to work well for small datasets with many columns.


Consider the FastFDs \cite{fastfd} algorithm for FD discovery.  
Returning to Figure~\ref{fig:small2}, we get the following evidence sets (expressed concisely as attributes whose values are different) from the six tuple pairs:

\begin{table}[h!]
\centering
\resizebox{\textwidth}{!}{
\begin{tabular}{ |l| }
\hline
$EV(t_1, t_2)$ = $\{A, B\},$  $EV(t_2, t_3)$ = $\{A, B, D\},$ $EV(t_1, t_4)$ = $\{A, B, C, D\},$ \\
$EV(t_1, t_3)$ = $\{B, D\},$  $EV(t_2, t_4)$ = $\{B, C, D\},$  $EV(t_3, t_4)$ = $\{A, C, D\}$ \\ \hline
\end{tabular}
}
\end{table}


After FastFDs generates evidence sets, for each possible right-hand-side attribute of an FD, it finds all the left-hand-side attribute combinations that hold.  Say $A$ is the right-hand side attribute currently under consideration. The algorithm first removes $A$ from the evidence sets, giving $\{\{B\}, \{B, C, D\}, \{B,D\}, \{C, D\}\}$. Next, FastFDs finds \emph{minimal covers} of this set, i.e., minimal sets of attributes that intersect with every evidence set.  In this example, we get $BC$ and $BD$, and therefore we conclude that $BC \rightarrow A$ and $BD \rightarrow A$.  FastDC \cite{fastdc} works similarly to discover DCs.


FastFDs avoids considering all $n(n-1)/2$ pairs of tuples when generating evidence sets.  Instead, it only considers pairs of tuples that belong to the same equivalence class for at least one attribute.  For example, in Figure~\ref{fig:small2}, tuples 1 and 4 are not in the same equivalence class for any of the four attributes.  In these cases, a tuple pair has no attributes in common and therefore the corresponding evidence set is all of $R$, which trivially intersects with every possible cover.


\begin{table*}[t]
\centering
\caption{Summary of primitives and their usage across algorithms}
\begin{tabular}{ |c|c|c|c|c|c|c|c| } 
 \hline
 Primitive & \textbf{TANE} & \textbf{FASTOD} & \textbf{FastFDs} & \textbf{FastDCs} & \textbf{HyFD} & \textbf{HyUCC} & \textbf{Hydra} \\\hline
 $genEQClass(X, I)$ & $\checked$ & $\checked$ & $\checked$ & & $\checked$ & $\checked$ & $\checked$ \\\hline
 $genEVSet(t_i, t_j, P)$ &  &  & $\checked$ & $\checked$ & $\checked$ & $\checked$ & $\checked$ \\\hline
 $checkRefinement(X, Y, I)$ & $\checked$ & $\checked$ & & & $\checked$ & $\checked$ &  \\\hline
 $join(S_i, S_j, p)$ & $\checked$ & $\checked$ & $\checked$ & $\checked$ & $\checked$ & $\checked$ & $\checked$ \\\hline
 $setCover(S)$ &  &  & $\checked$ & $\checked$ & $\checked$ & $\checked$ & $\checked$ \\\hline
 $sort(S, Comparator)$ &  & $\checked$ & $\checked$ & $\checked$ & $\checked$ & $\checked$ & $\checked$ \\\hline
\end{tabular}
\label{tab:primitives}
\end{table*}

\textbf{Hybrid algorithms} These algorithms switch back and forth between schema-driven and data-driven phases; examples include HyFD \cite{hyfd} for FDs, HyUCC \cite{hyucc} for UCCs and Hydra \cite{hydra} for DCs.  For example, HyFD starts with a data-driven phase, but it generates evidence sets only from a \emph{sample} of tuples.  
From these evidence sets, HyFD identifies potential FDs, which are those that have not been violated by the sampled tuple pairs (but may be violated by some other tuple pairs).
To represent these potential FDs, HyFD uses an \textit{FDTree} data structure, which is a prefix-tree, each of whose nodes corresponds to a set of attributes. 
Next, to validate the potential FDs, HyFD switches to a schema-driven phase, which traverses the FDTree level-wise, similarly to how TANE traverses the attribute lattice level-wise.
At some point, HyFD may switch back to a data-driven phase and generate evidence sets from a different sample of tuples, and so on.
HyUCC \cite{hyucc} (which discovers UCCs) and Hydra \cite{hydra} (which discovers DCs) are similar to HyFD but Hydra switches only once from the data-driven phase to the schema-driven phase.

%% file: primitives.tex
Our approach to design efficient distributed methods for dependency discovery is to identify the data processing steps and explore different implementations of these steps. 
To realize this approach, we propose a general framework consisting of six primitives listed below.
We identified the primitives by decomposing existing algorithms into common data processing steps whose distributed implementations are well-understood. For example, as explained in Section~\ref{sec:algos}, generating the next lattice level in schema-driven algorithms consists of a join (to generate new lattice nodes) and a group-by operation (to generate new equivalence classes). Similarly, generating evidence sets consists of a join.

\begin{enumerate}
\item \textbf{Generate equivalence classes ($genEQClass(X,I)$):} Given an attribute set $X \subseteq R$ and some input data $I$, this primitive computes $\pi_X$, i.e., it partitions $r$ according to $X$.
This is similar to the relational group-by operator and can be implemented by sorting or hashing the data.  
This primitive is used to verify if dependencies hold by schema-driven algorithms and to decide which tuple pairs to examine by data-driven algorithms. 

\item \textbf{Generate evidence set ($genEVSet(t_i, t_j, P)$):} Given a pair of tuples $t_i$ and $t_j$, from $r$, this primitive generates a set of dependencies (defined under a predicate space $P$) that are violated by this tuple pair, i.e. the evidence set of ($t_i$, $t_j$). 
Recall from Section \ref{sec:preliminaries} that DCs have the most general predicate space while FDs, ODs and UCCs have simpler predicate spaces. This primitive is used in data-driven and hybrid algorithms.   



\item \textbf{Partition refinement ($checkRefinement(X, Y, I)$):} Given two attribute sets $X, Y \subseteq R$, and some input data $I$, this primitive returns true if the partitioning of $r$ under $X$ ($\pi_X$) refines the partitioning of $r$ under $Y$ ($\pi_Y$) and false otherwise.  Schema-driven and hybrid algorithms use this primitive.  As discussed in Section~\ref{sec:algos}, in UCC and FD discovery, this primitive can return true or false by comparing the counts of $|\pi_X|$ and $|\pi_Y|$, whereas in OD discovery, it needs to check the ordering of tuples within matching equivalence classes.

\item \textbf{Join ($join(S_i, S_j,p)$):} This primitive joins two sets of elements, $S_i$ and $S_j$, using $p$ as the (optional) join predicate.  Schema-driven (and hybrid) algorithms use the join when generating attribute sets for the next lattice level; here, it is a self-join of the previous level's attribute sets.  Data-driven (and hybrid) algorithms join pairs of tuples to generate evidence sets.

\item \textbf{Cover ($setCover(S)$):} Given a set of evidence sets $S$, this primitive computes all minimal covers, and therefore the dependencies that hold given $S$.  Cover is used in all data-driven and hybrid algorithms.



\item \textbf{Sort ($sort$ $(S, Comparator)$):} This primitive sorts the set $S$ based on the provided comparator.  FastOD sorts tuples within equivalence classes and checks for proper ordering during the partition refinement check.  Data-driven algorithms sort evidence sets based on their cardinality to speed up the minimal cover operation.  Hybrid algorithms use sorting during sampling (of tuple pairs to generate evidence sets). 
\end{enumerate}

Table \ref{tab:primitives} highlights the expressiveness of the primitives and their usage across seven popular and state-of-the-art dependency discovery algorithms. 
As mentioned in Section~\ref{sec:intro}, we refer to logical rewrites of the algorithms using the primitives as \textit{logical discovery plans} (LDPs) and their physical implementations as \textit{physical discovery plans} (PDPs). 


\textbf{Design space:} There are many possible physical implementations of our primitives.  As in DBMSs, one important factor to consider is the size of the input compared to the available memory, e.g., to determine when to use a \textit{hash-join} or a \textit{sort-merge-join}. Similar choices exist in distributed frameworks such as Spark and Map-reduce. In particular, different distribution strategies have different memory footprints, suggesting a space of possible optimizations.
In Sections \ref{sec:tane}-\ref{sec:hyfd}, we explore this optimization space with the help of our primitives.  We consider TANE as a schema-driven example, FastFDs as a data-driven example and HyFD as a hybrid example (however, our conclusions apply to other algorithms within these three categories). For each case study, we show that different LDPs exist and we show two (of the many possible) distributed PDPs for each LDP.  One PDP, which we call \textit{large-memory plan} or lmPDP, assumes that each worker's memory is large enough for all computations; the other, which we call \textit{small-memory plan} or smPDP, assumes that the data may spill to external storage. For the PDPs, we assume Spark to be the data processing framework.
We show that different PDPs have different performance characteristics in terms of communication and computation cost, but we defer the issue of automatically selecting the best PDP for a given workload and system configuration to future work.


%% file: tane.tex

\begin{figure*}%
\centering
    \begin{subfigure}[c]{0.4\textwidth}
        \myalgorithma
        \caption{Logical discovery plan 1}
        \label{fig:tane-dist}
    \end{subfigure}\hspace{20pt}
    ~
    \begin{subfigure}[c]{0.4\textwidth}
        \myalgorithmb
        \caption{Logical discovery plan 2}
        \label{fig:tane-imp-dist}
    \end{subfigure}
    
\caption{TANE algorithm}
\end{figure*}

We start by studying TANE \cite{tane}.  As explained in Section \ref{sec:algos}, for each lattice level, schema-driven algorithms such as TANE compute dependencies holding in this level, prune the search space based on the discovered dependencies, and generate the next level.
The execution plans presented in this section can be easily extended to discover order dependencies \cite{fastod} as follows. Discovering ODs is more expensive than FDs because, as discussed in Section~\ref{sec:algos}, verifying ODs requires a refinement check and an ordering check.  Thus, to implement the $checkRefinement$ primitive, complete equivalence classes must be examined (not just the number of equivalence classes, i.e., the number of distinct values).

\subsection{LDP 1: Original TANE}
\label{sec:dist-tane}
Figure~\ref{fig:tane-dist} shows the first LDP written using our primitives, which follows the design principles of the original TANE algorithm: compute new equivalence classes by intersecting the previous level's equivalence classes.  To implement this, the input to the $genEQClass$ primitive in line 16 consists of a new attribute combination $X \cup Y$ and the equivalence classes $\pi_X$ and $\pi_Y$ from the previous level. Furthermore, the lattice stores attribute combinations and their associated equivalence classes (lines 3 and 17). 


\subsubsection{Large-memory PDP}
\label{sec:dist-tane-phy1}


\textit{Generating first level}: 
In lines 2-3, we generate equivalence classes for the first lattice level, i.e., for single attributes. This is implemented by distributing the columns in $R$ across the $k$ workers in a round-robin fashion.  Each worker scans the tuples in $r$ and hashes their values to compute equivalence classes for the columns assigned to it.

\textit{Computing dependencies}: 
As discussed in Section~\ref{sec:algos}, to check if an FD $X \setminus A \rightarrow A$ holds, it suffices to verify that $|\pi_{X \setminus A}| = |\pi_{X}|$. Thus, we distribute the counts of the equivalence classes (i.e. $|\pi_X|$) in the current lattice level $L_i$ (i.e., the dependencies to check) across the $k$ workers in a round-robin fashion and we broadcast the counts for attribute combinations in $L_{i-1}$ to each worker.

\textit{Pruning}: The driver machine then receives the discovered dependencies from the workers and applies pruning rules to the current lattice level.  Lattice nodes that have not been pruned are used to generate the next level. 

\textit{Generating next level}: 
This requires a (self-) join to produce new attribute combinations and their equivalence classes.  We implement the join as a map-reduce job, in which each worker creates a subset of nodes in the next lattice level.  We use a distributed self-join strategy called the \emph{triangle distribution strategy} \cite{xu}, which was shown to be optimal in terms of communication and computation costs. 
Next, a map job generates new equivalence classes, in which each worker creates equivalence classes for the nodes it has generated during the join.  New equivalence classes are created by intersecting pairs of equivalence classes from the previous level.  For example, if a worker receives equivalence classes for $AB$ and $AC$ during the join, it can create equivalence classes for $ABC$.  The new equivalence classes are written to the distributed filesystem.

\textbf{Cost analysis:}
To generate equivalence classes in the first lattice level, the computation is linear in terms of the number of tuples (single pass to hash the tuple values), and the columns are equally shared across all the workers.  The cost of computing dependencies is negligible since we only need to compare counts of equivalence classes (line 13).  
To generate the next level of equivalence classes, the cost of the triangle distribution strategy is given by \cite{xu}: $X_i \leq |I|*\sqrt{2/k}$, and $Y_i \leq |I|(|I|-1)/2k$, where $|I|$ is size of the input to the join, which is a lattice level ($L_{l}$) in our case. 
Due to pruning, we compute equivalence classes for $|L_{l+1}|$ attribute combinations and not for all pairs of attribute combinations resulting from the self-join. This approximation gives: $Y_i \leq 2n|L_{l+1}|/k$. Aggregating the costs for up to $m$ levels in the lattice, we get \textbf{X} $\leq 2^m n \sqrt{2/k}$, and \textbf{Y} $\leq 2^m 2n/k$. The factor of $2n$ in \textbf{Y} is because the intersection of two equivalence classes requires a scan over each one, which is $2n$ in the worst case.

\subsubsection{Small-memory PDP}
\label{sec:dist-tane-phy2}

Note that generating equivalence classes is memory-intensive due to the size of the attribute lattice. In the small-memory PDP, we focus on an alternative implementation of this task. 

\textit{Generating first level}: Our strategy in the lmPDP was to use each worker to generate equivalence classes for multiple columns. Here, our strategy is to use multiple workers to generate one column's equivalence classes. We do this by using Spark's distributed \textit{groupBy} operation. This has two advantages: it reduces the memory load per worker and allows Spark to take care of spilling the computation to disk.

\textit{Generating next level}: 
The lmPDP required $O(n2^m/\sqrt{k})$ memory for the triangle distribution strategy (which is the memory footprint given in \cite{xu}), and the equivalence classes assigned to a worker also must fit in its memory. In smPDP, we consider two regimes of memory capacity at a worker: (1) not enough memory to use the triangle distribution strategy, and (2) even less memory such that even the equivalence classes do not fit.  For regime (1), we implement the self-join using Spark's \textit{cartesian} operation and let Spark do the memory management. For regime (2), each worker reads the required equivalence classes $\pi_X$ and $\pi_Y$ from external storage in chunks (small enough to fit in memory) to create $\pi_{X \cup Y}$. Note that while doing this, a worker needs to make multiple passes over the input equivalence classes.

\textbf{Cost analysis:} To generate the first level's equivalence classes, the input to each \textit{groupBy} call is a column from $R$ of size $n$. In the worst case, the data can be skewed such that all $n$ tuples belong to the same group and are shuffled to a single worker, and this can happen for each column. Since each worker initially does roughly the same amount of computation in the map stage, we get $\textbf{X} \leq nm$ and $\textbf{Y} \leq nm/k$. The communication cost is greater than that for lmPDP for generating the first level. To generate the next level, the cost for both memory regimes is higher than the cost in lmPDP. For regime (1), Spark's \textit{cartesian} operation does more data shuffling than the triangle strategy. For regime (2), the cost is even greater due to the need to make multiple passes over the equivalence classes.

\subsection{LDP2: Modified TANE}
\label{sec:imp-tane}

LDP1 computes new equivalence classes by intersecting pairs of equivalence classes from the previous lattice level.  
This requires materializing and communicating equivalence classes to workers, which is expensive. In fact, the equivalence classes for a given lattice level may be larger than the input dataset.  
We now suggest an alternative LDP that \emph{computes new equivalence classes directly from the data instead of computing them using the previous level's equivalence classes}. 

Figure \ref{fig:tane-imp-dist} shows LDP2, with changes highlighted in blue. The primitive $genEQClass$ now takes as input a column combination and the tuples in $r$ (Line 3 and 16). Also, note the difference in line 3 and 17: a lattice level now includes attribute combinations and the number of the corresponding equivalence classes, not the equivalence classes themselves.



\subsubsection{Large-memory PDP}
\label{sec:imp-tane-phy1}
The implementation to generate equivalence classes for the first level and to compute dependencies is the same as in LDP1 from Section \ref{sec:dist-tane-phy1}. To generate the next level, we again use the triangle strategy to implement the self-join, which divides new attribute combinations among  workers.  Equivalence classes are not materialized with the corresponding attribute combinations, but we do need to store the number of equivalence classes for each attribute combination, which is needed to $checkRefinement$. 
Workers then compute the new equivalence classes assigned to them directly from the data (using hashing).

\textbf{Cost analysis:}
The cost to generate the first lattice level and compute dependencies is the same as in Section \ref{sec:dist-tane-phy1}. 
The cost of the self-join is negligible because it only involves attribute combinations and not the equivalence classes. Therefore it does not depend on the number of tuples $n$. 
To generate equivalence classes for a new lattice level, each worker is responsible for roughly the same number of attribute combinations (i.e. $|L_{l+1}|/k$), and using hashing it requires a single pass over the data (i.e., $nm$). Doing this for up to $m$ lattice levels gives: \textbf{Y} $\leq nm2^m/k$, and \textbf{X} $\leq nm*m$.
When compared to lmPDP in Section \ref{sec:dist-tane-phy1}, the communication cost of this plan is significantly lower because it avoids communicating the previous level's equivalence classes to the workers. 

Furthermore, LDP2 creates opportunities for further reduction of  communication cost for the Spark framework. The \textit{Broadcast} mechanism in Spark allows data to be cached at the workers for the lifetime of a job.  Thus, if each worker's memory is large enough to store the input dataset, it only needs to be sent once and can be reused for each lattice level. In our experimental evaluation (Section \ref{sec:eval}), we exploit this optimization whenever possible.

\subsubsection{Small-memory PDP}
\label{sec:imp-tane-phy2}

We borrow the strategy from Section \ref{sec:dist-tane-phy2}: we use multiple workers to generate equivalence classes for a particular attribute combination using Spark's \textit{groupBy}. The difference is that we only save the number of equivalence classes, not the equivalence classes themselves, since new equivalence classes are always computed from the input dataset, not from previous level's equivalence classes.

\textbf{Cost analysis:}
The cost of generating the first level is the same as in Section \ref{sec:dist-tane-phy2}, and the cost of the self-join and computing dependencies is the same as in Section \ref{sec:imp-tane-phy1}. To generate equivalence classes using \textit{groupBy} for a particular lattice level, the number of calls made to \textit{groupBy} is the same as the number of nodes in the lattice. In the worst case, each call will re-partition the data (i.e. $n$ tuples) to compute \textit{groupBy}. Hence, the cost of this smPDP is higher than the cost of the lmPDP (Section \ref{sec:imp-tane-phy1}) because it requires much more data shuffling. 
However, this smPDP still has a lower cost than the smPDP in Section \ref{sec:dist-tane-phy2} because it avoids the expensive \textit{cartesian} operation to create new equivalence classes by intersecting the previous level's equivalence classes (which, in the worst case, means that each worker may need to be sent the entire previous lattice level). 

%% file: fastfd.tex

\begin{figure}
\centering
    \begin{subfigure}{\textwidth}
        \myalgorithmc
        \caption{Logical discovery plan 1}
        \label{fig:fastfd-dist}
    \end{subfigure}\\ 
    \begin{subfigure}{\textwidth}
        \myalgorithmd
        \caption{Logical discovery plan 2}
        \label{fig:fastfd-imp-dist}
    \end{subfigure}
    
\caption{FastFDs algorithm}
\end{figure}

We now study FastFDs \cite{fastfd}. As explained in Section \ref{sec:algos}, the general strategy of data-driven algorithms is to generate evidence sets and then find minimal covers of the evidence sets.
We note that the plans shown in this section can also be applied to the FastDCs algorithm for discovering DCs.  The difference is that a richer predicate space will be used by $genEVSet$.   

\subsection{LDP1: Original FastFDs}
\label{sec:dist-fastfd}
Figure \ref{fig:fastfd-dist} shows the first LDP that follows the main idea of original FastFDs algorithm \cite{fastfd}, which is to compare only those tuple pairs which belong to the same equivalence class for at least one attribute. Lines 8-9 compute the equivalence classes for all attributes in $R$. Then, lines 10-12 perform a join operation on each  equivalence class to compare tuples and generate evidence sets. Note that the predicate space used by $genEVSet$ consists of just inequalities (line 12), which is sufficient for FDs. Finally, we sort the evidence sets by their cardinality and compute their minimal covers (lines 4-5).


\subsubsection{Large-memory PDP}
\label{sec:dist-fastfd-phy1}
Implementing lines 8-9 is similar to generating the first level of equivalence classes in TANE (Section \ref{sec:dist-tane-phy1}): by distributing the columns among workers in a round-robin fashion, with each worker generating equivalence classes for multiple columns using hashing.

Next, generating evidence sets (lines 10-12) requires two jobs. First, a map-reduce job implements a self-join that joins pairs of tuples within the same equivalence class.  For example, returning to Figure~\ref{fig:small2}, equivalence classes for $A$ generate tuple pairs (1,3) and (2,4); equivalence classes for $B$ generate (3,4), and so on. To implement this type of self-join in a distributed fashion, we use the \textit{Dis-Dedup$^+$} algorithm from \cite{xu}.  This algorithm was originally proposed for data deduplication, where a dataset is partitioned into blocks, potentially by multiple partitioning functions, and tuple pairs from the same block are checked for similarity.  Observe that our scenario is similar, in which a dataset is partitioned into blocks via equivalence classes and FastFDs only needs to compare tuple pairs from the same equivalence class (block).  Afterward, a map job generates evidence sets, in which each worker computes evidence sets for the tuple pairs it created during the self-join.

Finally, we sort the equivalence classes and compute minimal covers.  We do these steps locally at the driver node because FastFDs uses a depth-first-search strategy to compute all minimal covers, which is inherently sequential \cite{makki, reif}.


\textbf{Cost analysis:} 
The cost analysis for generating equivalence classes is the same as for  TANE in Section \ref{sec:dist-tane-phy1}. Next, we examine the cost of generating evidence sets. If the size of an equivalence class $j$ is $B_j$, then the number of comparisons done to generate evidence sets for all tuple pairs from this equivalence class is $B_j(B_j-1)/2 \approx B_j^2/2$. Assuming $c$ is the total number of equivalence classes, the total number of comparisons when generating evidence sets is $W = \sum_{j=1}^{c}B_j^2/2$. Each tuple pair comparison takes $m$ amount of work, therefore the total work done is $m*W$. With this, we can directly use the cost analysis of \textit{Dis-Dedup$^+$} from \cite{xu}, which gives us \textbf{X} $\leq 5m^2 max(n/k, \sqrt{2W/k})$, and \textbf{Y} $\leq 5mW/k$. Note that we have $m$ ``blocking functions'' and $m$ amount of work is required to compare (all $m$ attributes of) each tuple pair.


\subsubsection{Small-memory PDP}
\label{sec:dist-fastfd-phy2}

To generate equivalence classes (lines 8-9 in Figure \ref{fig:fastfd-dist}), we again use multiple workers to generate equivalence classes for one attribute using Spark’s \textit{groupBy}, as in Section \ref{sec:dist-tane-phy2} for TANE. To generate evidence sets, (lines 10-12) we can off-load the (self) join operation (line 11) to Spark using the \textit{cartesian} operation over RDDs (but we need to filter out redundant pairs). In this case, Spark internally does the memory management of the RDD, spilling to disk if required. Note that the  \textit{cartesian} operation will be called once for each equivalence class.

\textbf{Cost analysis}: The cost analysis of generating equivalence classes is same as in Section \ref{sec:dist-tane-phy2}. Next, each call to the  \textit{cartesian} operation shuffles tuples from the input equivalence class across multiple workers such that some tuples might be sent to multiple (or all) workers. When combining the data shuffle across all calls to the \textit{cartesian} operation, each worker might end up seeing close to all the input tuples multiple times. This gives a much higher communication cost compared to the lmPDP (Section \ref{sec:dist-fastfd-phy1}). 

\subsection{LDP2: Modified FastFDs}
\label{sec:imp-fastfd}

The \textit{Dis-Dedup$^+$} algorithm is the current state-of-the-art, but it still incurs a non-trivial communication and computation cost.  One problem is the redundant pair-wise tuple comparisons. Consider the equivalence classes $\pi_A = \{\{1,3\}, \{2,4\}\}$ and $\pi_C = \{\{1,2,3\}, \{4\}\}$ from the example in Section \ref{sec:algos}. According to LDP1, tuple 1 and tuple 3 are compared twice because they co-occur in two partially overlapping equivalence classes. Also, increasing the number of attributes increases the overlap of equivalence classes, thereby increasing the number of redundant pair-wise tuple comparisons.  This is also evident from the $m^2$ factor in the cost analysis of \textit{Dis-Dedup$^+$} in Section \ref{sec:dist-fastfd-phy1}. It is possible to eliminate this problem, but it would require an expensive comparison of all pairs of tuples in order to eliminate duplicate tuple pairs. In LDP2, we explore this idea, which trades off communication for computation. Figure \ref{fig:fastfd-imp-dist} shows the pseudocode for LDP2, with changes  highlighted in blue. In particular, in line 7, we perform a self $join$ on the complete relation $r$.

\subsubsection{Large-memory PDP}
\label{sec:imp-fastfd-phy1}
We again use the triangle join strategy from \cite{xu} to implement the $join$ in line 7. This requires one map-reduce job to compute a full self-join of $r$ and then a map job to generate evidence sets from all tuple pairs.


\textbf{Cost analysis:} Applying the cost analysis for triangle join strategy, we get: \textbf{X} $\leq nm\sqrt{2/k}$ and \textbf{Y} $\leq mn^2/2k$. This is an improvement over the cost of lmPDP in Section \ref{sec:dist-fastfd-phy1}, specially when $m$ is large, which is a typical use case for FastFDs.



\subsubsection{Small-memory PDP}
\label{sec:imp-fastfd-phy2}
The triangle join strategy in the lmPDP has a memory footprint of $O(nm/\sqrt{k})$ per worker. When each worker's memory is smaller than that, we off-load the $join$ implementation to Spark's \textit{cartesian} operation (we filter out redundant tuple pairs), and let Spark do the memory management.

\textbf{Cost analysis}: \cite{xu} showed that triangle strategy is optimal in terms of communication cost when implementing the self-join. Therefore, the cost of implementing the self-join using the \textit{cartesian} operation cannot be lower. However, compared to the smPDP in Section \ref{sec:dist-fastfd-phy2}, this implementation can still perform better when $m$ is large because each tuple is compared exactly once.

%% file: hyfd.tex
\begin{figure*}[t!]
    \centering
    \begin{subfigure}[t]{0.49\textwidth}
        \centering
        \includegraphics[height=10.1cm, width=7.5cm,valign=c]{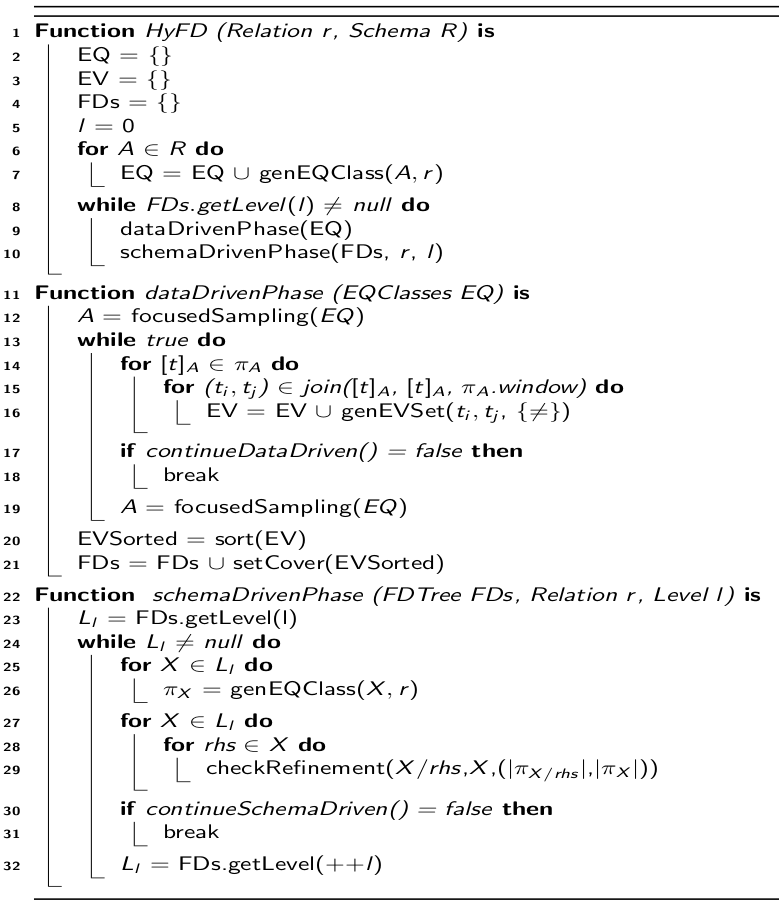}
        \caption{Logical discovery plan 1}
        \label{fig:hyfd-dist}
    \end{subfigure}
    ~
    \begin{subfigure}[t]{0.49\textwidth}
        \centering
        \includegraphics[height=10.1cm, width=7.5cm,valign=c]{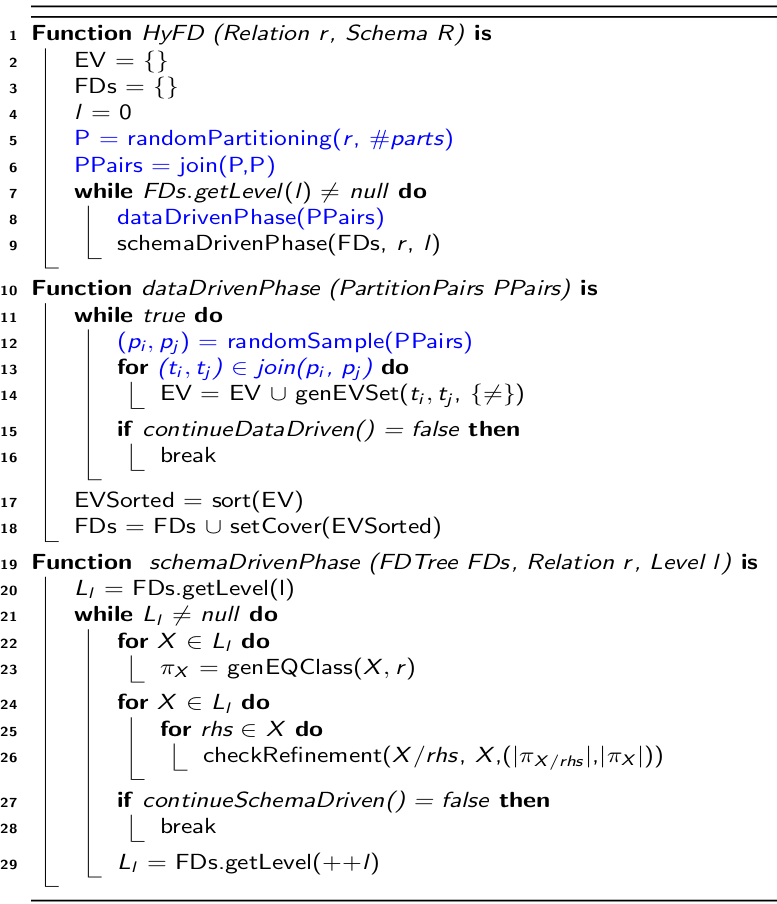}
        \caption{Logical discovery plan 2}
        \label{fig:hyfd-imp-dist}
    \end{subfigure}
    \caption{HyFD algorithm}
\end{figure*}

In this section, we analyze the HyFD algorithm \cite{hyfd}. As outlined in Section~\ref{sec:algos}, HyFD alternates between data-driven and schema-driven phases. 
We note that HyUCC \cite{hyucc} and Hydra \cite{hydra} can also be implemented using the plans described in this section, with some modifications. HyUCC \cite{hyucc} works similarly to HyFD, with pruning rules designed for UCC discovery. Hydra discovers DCs and hence uses a richer predicate space than FD and UCC discovery. Unlike HyFD, Hydra switches from the data-driven phase to the schema-driven phase only once, after the rate of generating DC violations drops below a user-supplied threshold.


\subsection{LDP1: Original HyFD}
\label{sec:dist-hyfd}
Figure~\ref{fig:hyfd-dist} shows the LDP of HyFD based on the original algorithm \cite{hyfd}. As in FastFDs, it begins by generating equivalence classes for all the attributes in $R$ (lines 6-7). Then in the data-driven phase, similar to FastFDs, it generates tuple pairs and the corresponding evidence sets. Compared to FastFDs, the difference is that not all tuple pairs are generated.  Instead, HyFD picks one attribute $A$ at a time and uses its equivalence classes to decide which tuple pairs should generate new evidence sets. To decide which attribute to use, the algorithm maintains a ranking of the attributes based on how many FDs have been violated according to their evidence sets.  This process is referred to in \cite{hyfd} as \emph{focused sampling} (line 12).

Next, two tuples, $t_i$ and $t_j$, within the same equivalence class are compared only if $j - i = window$, where $j$ and $i$ are their positions in the equivalence class, and $window$ is a threshold, with different attributes having different values of $window$.  This corresponds to a join with a $window$ predicate in line 15.  Whenever an attribute is selected in the data-driven phase, its $window$ value is incremented, which leads to new tuple pairs being generated.

The data-driven phase stops when newly generated evidence sets fail to identify new FD violations (encapsulated in the $continueDataDriven$ check in line 17). The evidence sets collected so far (line 16) are then used to generate FDs that have not yet been violated via set cover (lines 20-21) and inserts them into the FDTree.

The schema-driven phase traverses the FDTree level-wise; the $getLevel$ function in lines 23 and 32 retrieves all attribute sets from a given level.  For each attribute set, HyFD checks which FDs hold via $checkRefinement$ (Line 29).  The original HyFD implementation computes equivalence classes directly from the data (Line 26), and not by intersecting the previous level's equivalence classes.  This corresponds to our LDP2 for TANE (Section \ref{sec:imp-tane}). HyFD returns to the data-driven phase if the schema-driven phase spends too much time on a particular FDTree level (encapsulated in the $continueSchemaDriven$ function in line 30).  



\subsubsection{Large-memory PDP}
\label{sec:dist-hyfd-phy1}
We implement lines 6-7 in the same way we generated first-level equivalence classes in TANE in Section \ref{sec:dist-tane-phy1}.
Next, we use the following strategy to generate evidence sets in lines 14-16. If the selected attribute $A$ (in line 12) has $c$ equivalence classes, then we need to equally distribute these $c$ equivalence classes across $k$ workers.  As in \cite{xu}, we use a load balancing heuristic that arranges the equivalence classes in increasing order of their sizes, and divides them into $g = c/k$ groups, each group with $k$ equivalence classes.  Next, one equivalence class from each group is sent to a worker in round-robin fashion, such that each worker receives $g$ equivalence classes. Each worker then uses the $window$ parameter to decide which tuple pairs to generate. Finally, a map job generates evidence sets from the tuple pairs. The implementation of generating new equivalence classes and checking refinement is same as in TANE, described in Section \ref{sec:imp-tane-phy2}.

\textbf{Cost analysis:} The cost of generating equivalence classes is the same as in Section \ref{sec:dist-tane-phy1}. The cost of generating evidence sets in one iteration of the loop in lines 13-19 in the data-driven phase is \textbf{X} $\leq cm|B_{max}|/k$ and \textbf{Y} $\leq cm|B_{max}|/k$ where $B_{max}$ is size of the largest equivalence class (each worker receives $c/k$ equivalence classes and each of them could be of size at most $B_{max}$).

In the worst case, if HyFD discovers all the FDs using the data-driven phase, then the data-drive phase can be performed up to $n$ times (the size of the largest equivalence class for a given attribute can be $n$, and therefore the \emph{window} threshold can be incremented up to $n$ times). The cost of generating equivalence classes and checking refinement in the schema-driven phase is the same as in TANE in Section \ref{sec:imp-tane-phy1}.

\subsubsection{Small-memory PDP}
\label{sec:dist-hyfd-phy2}
In the lmPDP, we assigned multiple equivalence classes ($c/k$ of them) to each worker. In the smPDP, we reduce the memory footprint of each worker by assigning each equivalence class to multiple workers. Essentially, we use multiple workers to perform the $join$ in line 15 and then equally distribute all the generated pairs across workers to generate evidence sets. We implement the $join$ using Spark's \textit{join} operation and the $window$ parameter controls which keys to join. We borrow the implementation of generating equivalence classes (lines 6-7 and 25-26)  from TANE, as described in Section \ref{sec:imp-tane-phy2}

\textbf{Cost analysis}: The communication cost of this implementation is higher than that of lmPDP, simply because multiple rounds of data shuffle (one for each equivalence class) are needed to generate tuple pairs.
Additionally, the cost of generating equivalence classes in this PDP is higher than the cost in lmPDP in Section \ref{sec:dist-hyfd-phy1}. We know this from the cost analysis of TANE in Section \ref{sec:imp-tane-phy2}.

\subsection{LDP2: Modified HyFD}
\label{sec:imp-hyfd}

A drawback of LDP1 is its high communication cost during the data-driven phase, which is amplified if the data-driven phase is repeated multiple times. 
Also, as in FastFDs, there could be redundant evidence sets due to overlapping equivalence classes.

In LDP2 (Figure \ref{fig:hyfd-imp-dist}), we use the learnings from FastFDs: \emph{instead of computing evidence sets from tuple pairs that belong to the same equivalence class, we generate tuple pairs directly from the data}.  This means that focused sampling no longer applies as we are sampling tuples directly from the data and not from the equivalence classes over specific attributes.  We use random sampling without replacement, as explained below, which is easier to parallelize.  As before, changes are highlighted in blue. In line 5, we randomly partition the dataset into $k$ groups, and in line 6, we generate all possible pairs of groups, including pairs of the same group. Then, the data-driven phase uses pairs of groups instead of equivalence classes.  In particular, line 12 samples $k$ pairs of groups \emph{without replacement}, and lines 13-14 join each pair of groups to generate tuples pairs and the corresponding evidence sets.




\subsubsection{Large-memory PDP}
\label{sec:imp-hyfd-phy1}
Random partitioning of the data in line 5 is implemented using a simple map-reduce job: mappers assign partition IDs to tuples and reducers group tuples belonging to the same partition ID. Then, another map-reduce job implements the $join$ in Line 6 which generates pairs of groups. 

In line 12, we sample $k$ pairs of groups \emph{without replacement} which are distributed across $k$ workers. Therefore, each worker is responsible for generating evidence sets from one pair of groups. The implementation of equivalence class generation and checking refinement is the same as in LDP1 in Section \ref{sec:dist-hyfd-phy1}.

\textbf{Cost analysis:}
With $k$ workers, the cost of generating evidence sets in each iteration of the data-driven phase is: \textbf{X} $\leq 2nm/k$ and \textbf{Y} $\leq mn^2/k^2$. This is because two groups of size $mn/k$ each are sent to every worker and every worker generates all the tuple pairs, i.e., $mn^2/k^2$. The data-driven phase can run up to $(k+1)/2$ times.  To see this, note that every group must be joined with every other group, including itself, which amounts to $k(k+1)/2$ group-wise comparisons. With $k$ workers in parallel, each working on one group-pair, the $k(k+1)/2$ comparisons can be packed into $(k+1)/2$ jobs. This gives \textbf{X} $\leq (2nm/k)*(k+1)/2 \approx nm$, which is the size of the data. Compared to the lmPDP in Section \ref{sec:dist-hyfd-phy1}, this implementation performs more computation but much less data communication. The cost of the schema-driven phase is the same as in TANE in Section~\ref{sec:imp-tane-phy1}. The random partitioning step in line 5 and the join in line 6 use tuple identifiers and do not involve significant computation (only a linear scan of tuple IDs), and hence their cost is negligible compared to the other operations.

In this LDP (and its lmPDP and smPDP), we use a cost-based approach to decide when to switch between the phases. That is, we switch to the other phase if the communication cost plus the computation cost of proceeding with the current phase exceeds the cost of operating in the other phase.
The switching conditions are encapsulated in the $continueDataDriven$ (line 15) and the $continueSchemaDriven$ (line 27).

\subsubsection{Small-memory PDP}
\label{sec:imp-hyfd-phy2}
The data-driven phase of lmPDP assumes that two groups of size $nm/k$ each fit in each worker's memory to generate evidence sets. However, we can reduce the size of each group by creating more groups in line 5. The drawback is that we will perform fewer comparisons in each round of the data-driven phase, hence possibly generating fewer evidence sets. 
The implementation of generating  equivalence  classes  (lines  6-7 and 25-26) is borrowed from TANE, as described in Section \ref{sec:imp-tane-phy2}.

\textbf{Cost analysis}: The cost of generating evidence sets in each iteration of the data-driven phase will reduce in proportion to the number of groups generated in line 5. However, the data-driven phase can run more times as compared to the lmPDP (Section~\ref{sec:imp-hyfd-phy1}). For example, if we generate $2k$ groups instead of $k$ groups (as in Section~\ref{sec:imp-hyfd-phy1}), then the data-driven phase can run up to $(2k + 1)$ times.  With $k$ workers in parallel, each working on one group-pair, the $2k(2k+1)/2$ comparisons can be packed into $(2k+1)$ jobs. This gives \textbf{X} $\leq (2nm/2k)*(2k+1) \approx 2nm$, which is twice the communication cost in Section~\ref{sec:imp-hyfd-phy1}. Additionally, the cost of generating equivalence classes in this PDP is higher than the cost in Section~\ref{sec:imp-hyfd-phy1}. We know this from the cost analysis for TANE in Section \ref{sec:imp-tane-phy2}.

%% file: evaluation.tex
We now present our empirical evaluation. In Section~\ref{sec:in-depth} we first show that the large memory PDPs of the modified logical plans (LDP2s) are more computation and communication efficient than the large memory PDPs of LPD1s.
Next, in Section~\ref{sec:small-memory}, we show that our smPDPs can discover dependencies when worker memory is small compared to the data size, but their runtimes are significantly higher. We then demonstrate that the large memory PDPs of LPD2s can scale well with the number of worker machines, rows, and columns (Section~\ref{sec:scalability}). Finally, we examine the relative performance of various algorithms on different datasets (Section~\ref{sec:diff-datasets}).


\subsection{Experimental Setup}

We performed the experiments on a 6-node Spark 2.1.0 cluster.  Five machines run Spark workers and one machine runs the Spark driver. Each worker machine has 64GB of RAM and 12 CPU cores and runs Ubuntu 14.04.3 LTS.  On each worker machine, we spawn 11 Spark workers, each with 1 core and 5GB of memory.  The driver machine has 256GB of RAM and 64 CPU cores, and also runs Ubuntu 14.04.3 LTS.  The Spark driver uses one core and 50GB of memory.  We run Spark jobs in standalone mode with a total of 55 executors (11 workers times 5 worker nodes).

All algorithms are implemented in Java.  We obtained the source code for TANE, FastFDs, HyFD, Hydra, and HyUCC from the Metanome GitHub page \cite{metanome-git}.  We obtained the source code for FastDCs and FastODs from the respective authors. 
We use similar datasets as those used in recent work on dependency discovery \cite{pe2015}.
Their properties 
are summarized in Table \ref{tab:large-datasets}.  For reproducibility, all the algorithms, testing scripts and links to the datasets are publicly available on our GitHub page\footnote{https://github.com/hemant271990/distributed-dependency-discovery}.



\subsection{Comparison of LPD1s and LPD2s}
\label{sec:in-depth}

We use two datasets to compare the two LDPs studied in Sections \ref{sec:tane}-\ref{sec:hyfd}: one with a large number of rows (\textit{lineitem}) and one with a large number of columns (\textit{homicide}).  To ensure that the LDP1 implementations finish within a reasonable time, we delete a fraction of rows from these datasets.  
For \textit{lineitem}, we use 0.5 million rows, and for \textit{homicide}, we use 100,000 rows. We focus on large memory PDPs for this comparison because from our analysis we know that the runtime in the small memory regime is significantly higher (also shown empirically in Section \ref{sec:small-memory}). Therefore, for both LDPs we use the lmPDP as described in Sections \ref{sec:tane}-\ref{sec:hyfd}.
For each tested algorithm, we measure communication costs and runtime. 

\begin{table}[t]
\centering
\resizebox{\textwidth}{!}{
\caption{Runtime and data shuffle costs of TANE}
\label{tab:tane1}
\begin{tabular}{ c|c|c }
\hline
\textbf{lineitem 0.5Mx16} & Total time (secs) & Total shuffle (MB) \\ \hline
LDP1	& 722 & 355.8 \\ 
LDP2 & 100 & 24 \\ \hline 
\textbf{homicide 100Kx24} & Total time (secs) & Total shuffle (MB) \\ \hline
LDP1	& TLE & TLE \\ 
LDP2 & 23853 & 2.2 \\ 
\end{tabular}
}
\end{table}

\textbf{TANE}: Table \ref{tab:tane1} shows the runtime and maximum data sent (shuffled) to any worker for TANE LDP1 (Section~\ref{sec:dist-tane}) and LDP2 (Section~\ref{sec:imp-tane}). LDP1 exceeded the time limit of 24 hours (denoted ``TLE'') on the \textit{homicide} dataset and was nearly an order of magnitude slower on the \textit{lineitem} dataset (due to high data communication). For LDP2, we cache the dataset at the workers (using Spark's \textit{Broadcast} mechanism) to avoid re-reading it when computing equivalence classes for the next level.  


\begin{table}[t]
\centering
\resizebox{\textwidth}{!}{
\begin{tabular}{ c|c|c }
\hline
\textbf{lineitem 0.5Mx16} & Total time (secs) & Total shuffle (MB) \\ \hline
LDP1	& 5242 & 22.8 \\ 
LDP2 & 2098 & 5.4 \\ \hline 

\textbf{homicide 100Kx24} & Total time (secs) & Total shuffle (MB) \\ \hline
LDP1	& 1556 & 7.4 \\ 
LDP2 & 106 & 0.9 \\ 
\end{tabular}
}
\caption{Runtime and data shuffle costs of FastFDs}
\label{tab:fastfd1}
\end{table}

\textbf{FastFDs}: Table \ref{tab:fastfd1} shows the runtime and maximum data sent to any worker for FastFDs LDP1 (Section~\ref{sec:dist-fastfd}) and LDP2 (Section~\ref{sec:imp-fastfd}).  Again, LDP2 is significantly more time and communication-efficient. 
In Section \ref{sec:fastfd}, we pointed out that the computation and communication cost of the LDP1 becomes worse as the number of attributes increases.  This is evident from Table \ref{tab:fastfd1}, where the improvement on \textit{homicide} is 14x and on \textit{lineitem} it is 2.5x.

\textbf{HyFD}: Table \ref{tab:hyfd1} shows the runtime and maximum data sent to any worker for HyFD LDP1 (Section~\ref{sec:dist-hyfd}) and LDP2  (Section~\ref{sec:imp-hyfd}).  We break down the costs between the data-driven phase and the schema-driven phase.  As discussed in Section~\ref{sec:hyfd}, the data-driven phase can lead to a high volume of data shuffle if the dataset has many columns.  This explains why LDP1 performs poorly on \textit{homicide} which has 24 columns. Due to the large schema of \textit{homicide}, LDP2 spent most of the time in the data-driven phase. On \textit{lineitem}, the algorithm spends more time in the schema-driven phase because of the smaller schema.  Here, LDP2 started with data-driven phase but immediately switched to the schema-driven phase after one round and spent the rest of the time there because the cost of comparing 9090*9090 tuple pairs (500,000/55 = 9090) is higher than validating even the largest level of the FD tree.  In other words, LDP2 acted like TANE.  On the other hand, LDP1 did a few rounds of sampling, and due to the focused sampling strategy, it was able to discover a significant number of non-FDs.  This significantly pruned the search space of the schema-driven phase, allowing LDP1 to be as fast as LDP2.  In LDP2, the data shuffle cost is roughly equivalent to the size of the dataset because the broadcast data is cached in the workers' memory and therefore we only need to send it once at the beginning of the data-driven phase.

\begin{table*}[t]
\centering
\resizebox{\textwidth}{!}{
\begin{tabular}{ c|cccc|ccc }
\hline
\textbf{lineitem 0.5Mx16} & data-driven (secs) & schema-driven (secs) & rest (secs) & \textbf{Total time} (secs) & data-driven (MB) & schema-driven (MB) & \textbf{Total shuffle (MB)} \\ \hline
LDP1	& 83 & 88 & 18 & 189 & 41 & 24 & 65 \\ 
LDP2 & 38 & 138 & 17 & 193 & 24  & - & 24\\ \hline 

\textbf{homicide 100Kx24} & data-driven (secs) & schema-driven (secs) & rest (secs) & \textbf{Total time} (secs) & data-driven (MB) & schema-driven (MB) & \textbf{Total shuffle (MB)} \\ \hline
LDP1	& 4675 & 139 & 15 & 4829 & 2553.8 & 2.2 & 2556.0 \\ 
LDP2 & 121 & 30 & 17 & 168 & 2.2 MB & - & 2.2 \\ 
\end{tabular}
}
\caption{Runtime and data shuffle costs of HyFD}
\label{tab:hyfd1}
\end{table*}

From Tables \ref{tab:tane1}, \ref{tab:fastfd1}, and \ref{tab:hyfd1}, we observe that HyFD LDP2 cannot beat TANE LDP2 on the \textit{lineitem} dataset and it cannot beat FastFDs LDP2 on the \textit{homicide} dataset. For the \textit{lineitem} dataset, although HyFD LDP2 spent almost all the time in the schema-driven phase, it could not finish before TANE LDP2 for two reasons. First, HyFD spends time creating partitions in the pre-processing step. Second, HyFD generates equivalence classes for slightly more attribute combinations than TANE because HyFD cannot prune the attribute combinations that are keys. We observed the second behaviour even in the original implementation of HyFD \cite{hyfd} by disabling the sampling phase. For the \textit{homicide} dataset, even though HyFD discovered all the FDs in the data-driven phase, it could not finish before FastFDs because the schema-driven phase was still executed to validate the discovered FDs. However, HyFD performs best when the dataset has a large schema and large number of rows, as shown by the \textit{homicide} and \textit{ncvoter} datasets in Table \ref{tab:large-datasets}.

\subsection{Comparison of smPDPs} 
\label{sec:small-memory}

\begin{table}[t]
\centering
\resizebox{0.85\textwidth}{!}{
\caption{Runtimes of smPDPs compared to lmPDPs (for LDP2)}
\label{tab:small-memory}
\begin{tabular}{ c|c|c|c }
\hline
\textbf{lineitem 6Mx16} & TANE & FastFDs & HyFD \\ \hline
lmPDP	& 1.9h & $\approx$3d & 3.9h \\ 
smPDP    & 3.9h & $\approx$106d & 8.5h \\
\end{tabular}
}
\end{table}

In this experiment, we reduce each worker's memory from 5GB to 1GB. This is small enough that for our largest dataset, \textit{lineitem}, its equivalence classes do not fit in a worker's memory. We use the smPDPs from our case studies (Sections \ref{sec:tane}-\ref{sec:hyfd}) which are designed for this scenario. We use LDP2 for each algorithm because in the previous experiment (Section \ref{sec:in-depth}) we saw that \textit{LDP1} has a higher runtime, which gets worse in the small memory regime.

Table \ref{tab:small-memory} shows the runtimes for TANE, FastFDs and HyFD. The runtime of the smPDPs is almost twice as high as the corresponding lmPDPs for TANE and HyFD. For FastFDs, we extrapolate the runtimes by running it on datasets with 100K, 300K, 500K, and 700K rows because FastFDs has quadratic complexity in number of rows, and \textit{lineitem} has 6 million rows. The smPDP of FastFDs is an order of magnitude slower than the lmPDP. 

%

%



\begin{figure}[t]
\centering
\includegraphics[height=5cm, width=8.5cm]{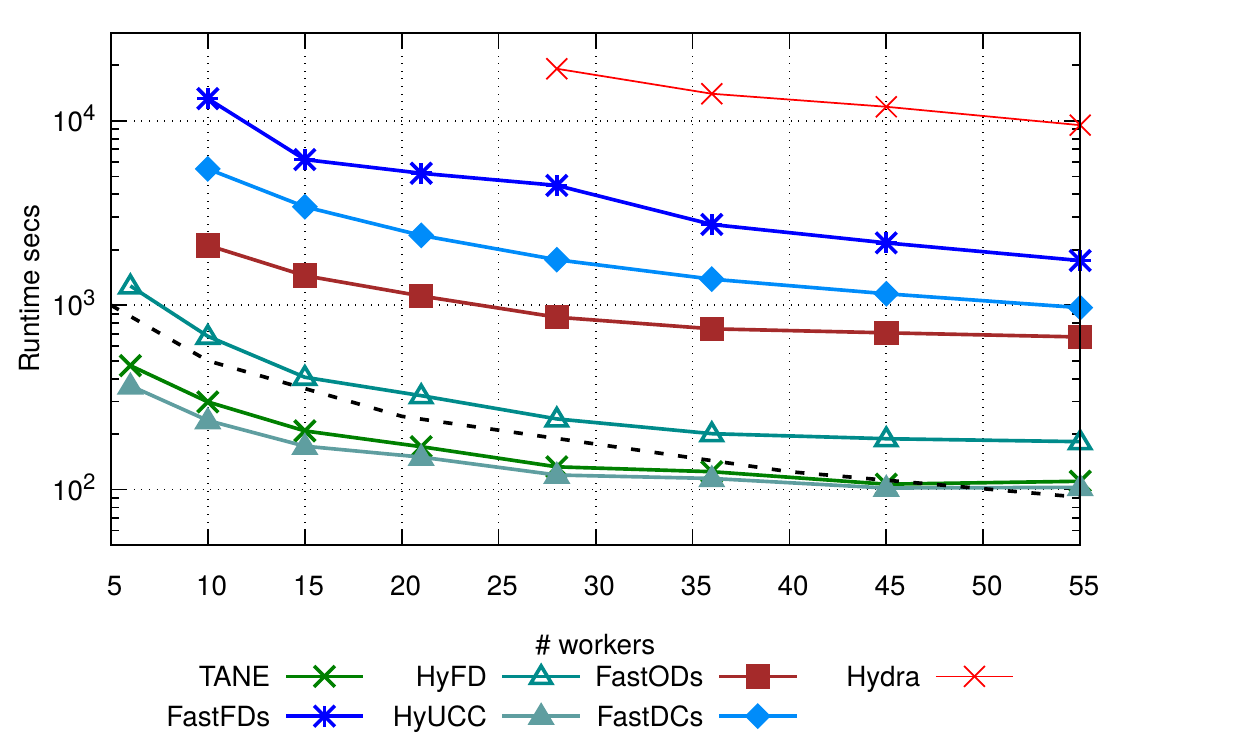}
\caption{Scalability with the number of workers (lmPDP of LDP2)}
\label{fig:workers}
\end{figure}

\begin{figure*}[h!]
    \centering
    \begin{subfigure}[t]{0.5\textwidth}
        \centering
        \includegraphics[height=5cm, width=8.2cm,valign=t]{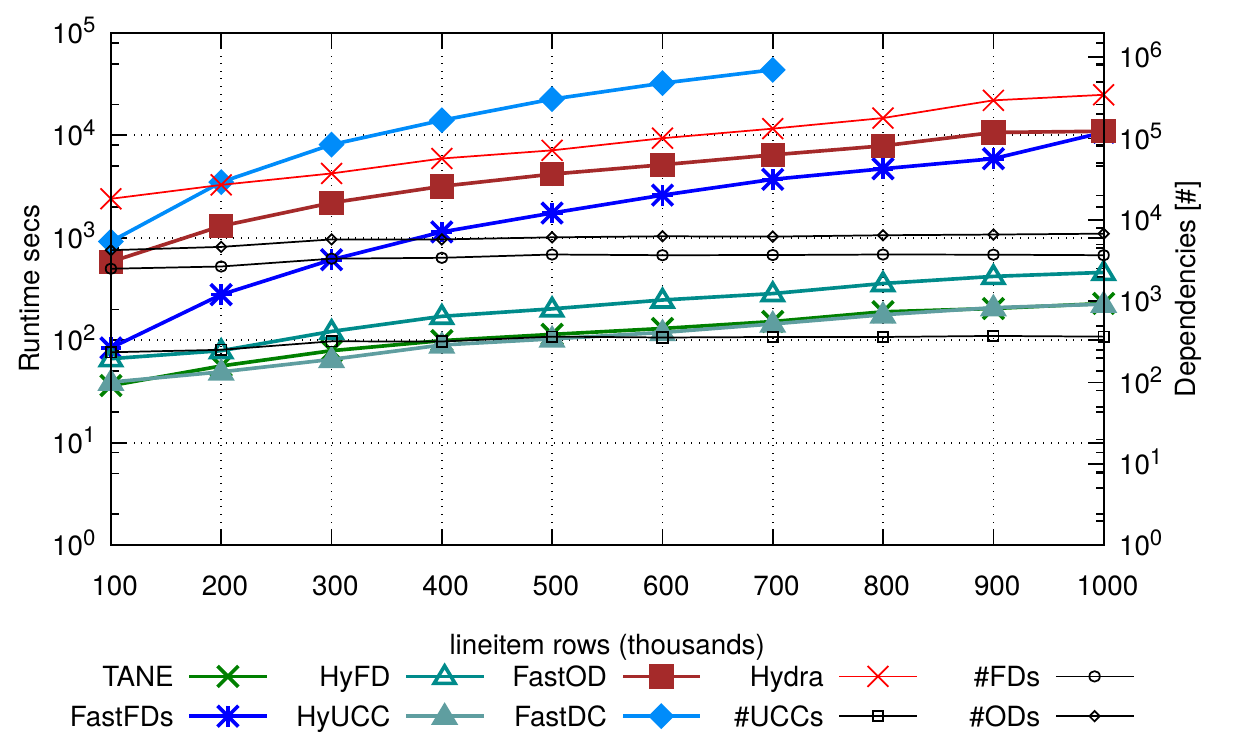}
    \end{subfigure}%
    ~ 
    \begin{subfigure}[t]{0.5\textwidth}
        \centering
        \includegraphics[height=5cm, width=8.2cm,valign=t]{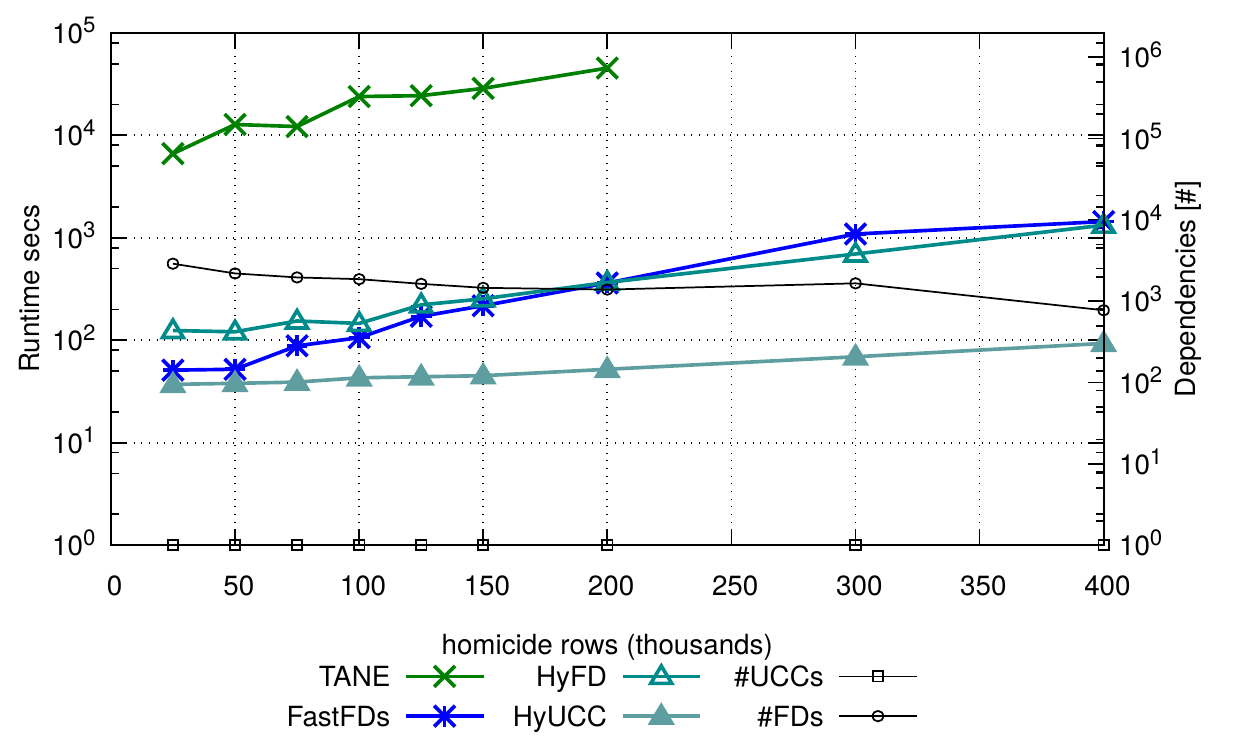}
    \end{subfigure}
    \caption{Scalability (of lmPDP of LDP2) with the number of rows (in thousands) for lineitem (left) and homicide (right). }
    \label{fig:rows}
\end{figure*}

\begin{figure*}[h!]
    \centering
    \begin{subfigure}[t]{0.5\textwidth}
        \centering
        \includegraphics[height=5cm, width=8.2cm,valign=t]{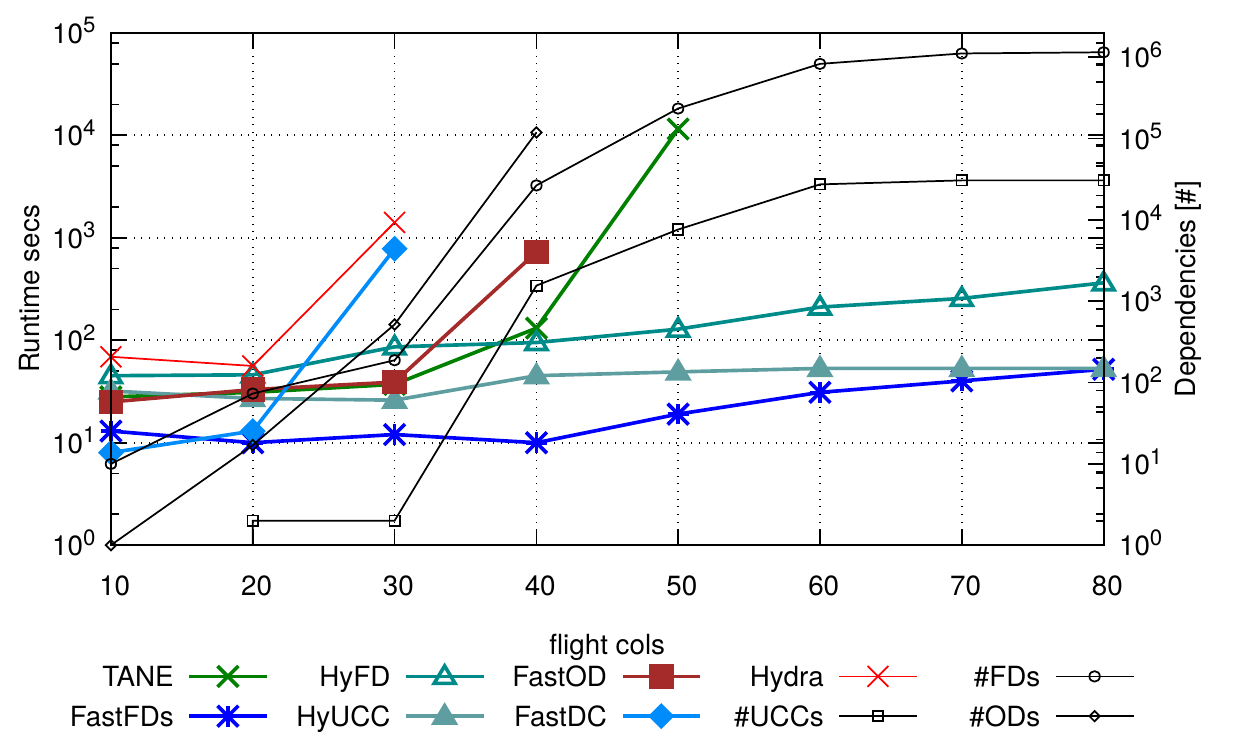}
    \end{subfigure}%
    ~ 
    \begin{subfigure}[t]{0.5\textwidth}
        \centering
        \includegraphics[height=5cm, width=8.2cm,valign=t]{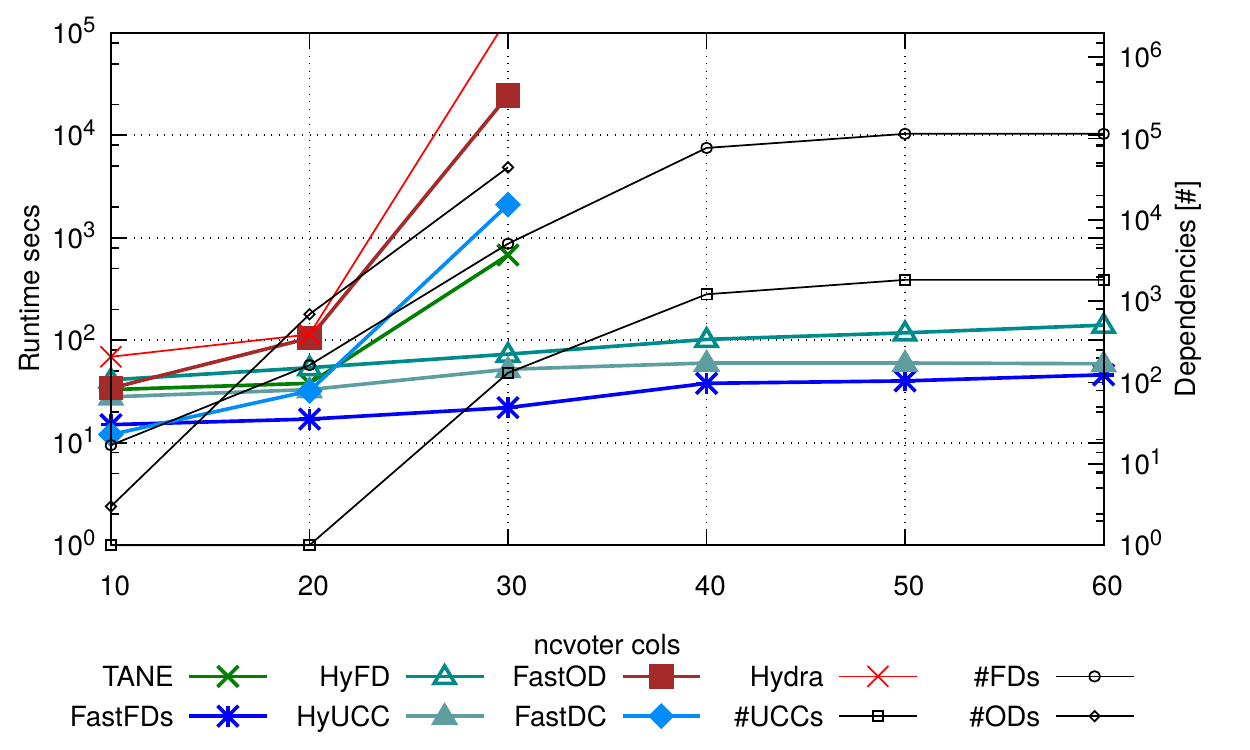}
    \end{subfigure}
    \caption{Scalability (of lmPDP of LDP2) with the number of columns for flight (left) and ncvoter (right)}
    \label{fig:cols}
\end{figure*}

\subsection{Scalability} \label{sec:scalability}

We now show the scalability of the distributed implementations of seven dependency discovery algorithms in terms of the number of workers, the number of rows, and the number of columns. In addition to TANE, FastFDs and HyFD, we also test FastODs, FastDCs, Hydra and HyUCC. We use lmPDP and LDP2 to make sure all algorithms terminate in reasonable time. 
For DC discovery, it has been reported in previous work that FastDCs is significantly slower than than Hydra \cite{hydra}, so we restrict FastDCs to discover DCs with at most 5 predicates. Note that our distributed implementations perform DC discovery from evidence sets centrally at the driver (by finding a minimal set cover). Therefore, this restriction has the same impact on the distributed and non-distributed implementations.


\subsubsection{Worker Scalability}
\label{sec:worker-scalability}

We first demonstrate near-linear scalability with the number of workers.  We use the \textit{lineitem} dataset with 0.5 million rows for TANE, FastFDs, HyFD, HyUCC, and Hydra and with only 100k rows for FastOD, and FastDCs (because of their high sensitivity to the number of tuples, as will be shown in Section \ref{sec:row-scalability}).
We vary the number of Spark workers from 6 to 55.

Figure \ref{fig:workers} shows the results. Note that the y-axis is scaled logarithmically, and the dashed line shows perfect linear scaling for reference. FastFDs and FastDCs are impacted more by the number of workers because their complexity is quadratic in the size of the input.  TANE outperforms FastFDs because the dataset has a small schema, and HyFD closely follows TANE because HyFD spent most of the time in the schema-driven phase. Scale-out of FastODs is similar to TANE, and scale-out of HyUCC and Hydra is similar to HyFD.
Recall that FastODs and FastDCs are running on a smaller dataset, so their runtimes are relatively low.

\subsubsection{Row Scalability}
\label{sec:row-scalability}

Next, we test scalability with the number of rows. We use two large datasets: \textit{lineitem} and \textit{homicide}.  On the \textit{lineietm} dataset, we tested the performance of all seven algorithms: TANE, FastFDs, HyFD, HyUCC, FastOD, FastDCs, and Hydra.
For \textit{homicide}, which has a larger schema, we did not run FastDCs, Hydra and FastOD because these algorithms take much longer to complete when the schema is large.  Results are shown in Figure \ref{fig:rows} (again, with logarithmic y-axes), including algorithm runtimes and the numbers of dependencies that were discovered.  

\textbf{\textit{lineitem:}} 
TANE and FastOD behave similarly and their runtime grows almost linearly with the number of rows.  FastOD is similar to TANE but partition refinement for order dependency discovery is more expensive, resulting in much longer runtimes for FastOD compared to TANE.  HyFD and HyUCC behave similarly and they closely follow the scalability of TANE; they both spend most of the time in the schema-driven phase due to smaller schema of \textit{lineitem}.  HyUCC is similar to HyFD, as described in Section~\ref{sec:imp-hyfd}.

FastFDs and FastDCs perform similarly and their runtime grows almost quadratically with the number of rows.  However, for \textit{lineitem}, there are 64 predicates that define the space of DCs.  Thus, the minimal set cover operation in FastDCs \cite{fastdc} is significantly more expensive. 
As expected, the performance of Hydra is significantly better than FastDCs, even when we restrict FastDCs to DCs with at most 5 predicates.

\textbf{\textit{homicide:}} Due to the large schema, TANE performs poorly, but it still scales linearly with the number of rows. Again, the runtime of FastFDs increases almost quadratically with the number of rows. HyFD spends most of its time in the data-driven phase and hence follows the performance of FastFDs.

\subsubsection{Column Scalability}
\label{sec:column-scalability}

We now evaluate scalability with the number of columns.  We use two datasets with a sufficient number of columns: \textit{flight} (with 1000 rows) and \textit{ncvoter} (with 10,000 rows).  We restrict FastDCs, Hydra and FastOD to fewer columns because of their high sensitivity to the schema size. Results are shown in Figure \ref{fig:cols}, including algorithm runtimes and the numbers of various dependencies that were discovered. Again, the y-axes are logarithmic.
As expected, TANE and FastODs runtimes increase exponentially with the number of columns because these algorithms are schema-driven.
FastDCs and Hydra runtimes increase exponentially because the predicates space of DCs increases significantly with the number of columns.
The runtime of FastFDs stays almost linear with the number of columns, and it performs best among the FD discovery algorithms.
HyFD performs similarl to FastFDs due to the low cost of the data-driven phase. However, HyFD still needs to switch to the schema-driven phase and hence it does not perform as well as FastFDs.  The behaviour of HyUCC is similar to HyFD. Recall that we restrict FastDCs to discover DCs only with up to 5 predicates, so its runtimes are lower than those of Hydra.

\subsection{Experiments on Different Datasets}
\label{sec:diff-datasets}
Finally, we evaluate (lmPDP implementations of LDP2s of) TANE, FastFDs and HyFD on several datasets with at least 14 and up to 109 columns. We omit FastDCs, Hydra and FastOD because these algorithms do not perform well on datasets with a large number of columns. Results are shown in Table \ref{tab:large-datasets}. \textit{Adult} is the smallest dataset, and all three algorithms finished in a reasonable time.  \textit{lineitem} has a large number of rows (6 million), meaning that FastFDs struggles but TANE and HyFD perform better.  
However, HyFD takes longer than TANE because, as mentioned before, HyFD incurs the overhead of creating partitions and it does not prune keys.
\textit{homicide} and \textit{ncvoter} are examples where HyFD switches between the two phases and discovers FDs the fastest.  For \textit{ncvoter}, FastFD ran out of memory at the driver because the search space for minimal set covers grew large.  For \textit{fd-reduced}, TANE performs best because almost all of the discovered FDs are present in the third level of the lattice; this is observed in HyFD \cite{hyfd} as well.
For the \textit{flight} dataset, HyFD spent most of the time in the data-driven phase, but it still had to validate millions of FDs in the schema-driven phase, and hence it could not beat FastFDs.

Recent work \cite{pe2015,hyfd} has compared FD discovery algorithms on similar datasets and concluded that schema-driven algorithms are suitable for datasets with many rows and data-driven algorithms are suitable for the datasets with many columns.  Hybrid algorithms perform best by spending most of their time in either the data-driven phase or the schema-driven phase, depending on their relative cost.  We observed similar trends in the distributed versions of these algorithms.


\begin{table}[h!]
\centering
\resizebox{0.96\textwidth}{!}{
\begin{tabular}{ c|c c c|c c c }
\textbf{Dataset} & \# Columns & \# Rows & \# FDs & TANE & FastFDs & HyFD \\ \hline
adult & 14 & 32,560 & 78 & 50s & \textbf{23s} & 101s \\
lineitem & 16 & 6,000,000 & 4,145 & \textbf{6,854s} & $>$48h & 14,311s \\
homicide & 24 & 600,000 & 637 & 137,279s & 3,152s & \textbf{3,113s} \\
fd-reduced & 39 & 250,000 & 89,571 & \textbf{86s} & 648s & 228s \\
ncvoter & 60 & 1,000,000 & 2,638,634 & $>$48h & MLE & \textbf{43.2h} \\
flight & 109 & 1,000 & 1,150,815 & $>$24h & \textbf{99s} & 351s \\
\end{tabular}
}
\caption{Runtimes on different datasets}
\label{tab:large-datasets}
\end{table}